\documentclass[twocolumn]{aastex631}

\usepackage{graphicx}	
\usepackage{amsmath}	
\usepackage{amssymb}

\received{\today}
\revised{\today}
\accepted{\today}

\submitjournal{ApJ}

\newcommand\bb[1]{\mbox{\boldmath{$#1$}}}
\newcommand\grad{\bb{\nabla}}
\newcommand\bcdot{\,\bb{\cdot}\,}
\newcommand\btimes{\,\bb{\times}\,}
\newcommand\rme{{\rm e}}
\newcommand\rmd{{\rm d}}
\newcommand\imag{{\rm i}}

\shorttitle{Synchrotron firehose instability}
\shortauthors{Zhdankin et al.}

\begin{document}

\title{{Synchrotron Firehose Instability}}

\correspondingauthor{Vladimir Zhdankin}
\email{vzhdankin@flatironinstitute.org}

\author{Vladimir Zhdankin}
\affiliation{Center for Computational Astrophysics, Flatiron Institute, 162 Fifth Avenue, New York, NY 10010, USA}

\author{Matthew W.~Kunz}
\affiliation{Department of Astrophysical Sciences, Princeton University, 4 Ivy Lane, Princeton, NJ 08544, USA}
\affiliation{Princeton Plasma Physics Laboratory, P.O.~Box 451, Princeton, NJ 08543, USA}

\author{Dmitri A.~Uzdensky}
\affiliation{Center for Integrated Plasma Studies, Department of Physics, 390 UCB, University of Colorado, Boulder, CO 80309, USA}

\begin{abstract}
We demonstrate using linear theory and particle-in-cell (PIC) simulations that a synchrotron-cooling collisionless plasma acquires pressure anisotropy and, if the plasma beta is sufficiently high, becomes unstable to the firehose instability, in a process that we dub the {\it synchrotron firehose instability} (SFHI). The SFHI channels free energy from the pressure anisotropy of the radiating, relativistic electrons (and/or positrons) into small-amplitude, kinetic-scale magnetic-field fluctuations, which pitch-angle scatter the particles and bring the plasma to a near-thermal state of marginal instability. The PIC simulations reveal a nonlinear cyclic evolution of firehose bursts interspersed by periods of stable cooling. We compare the SFHI for electron-positron and electron-ion plasmas. As a byproduct of the growing electron-firehose magnetic field fluctuations, magnetized ions gain a pressure anisotropy opposite to that of the electrons. If these ions are relativistically hot, we find that they also experience cooling due to collisionless thermal coupling with the electrons, which we argue is mediated by a secondary ion-cyclotron instability. We suggest that the SFHI may be activated in a number of astrophysical scenarios, such as within ejecta from black-hole accretion flows and relativistic jets, where the redistribution of energetic electrons from low to high pitch angles may cause transient bursts of radiation.
\end{abstract}

\keywords{plasma astrophysics, high-energy astrophysics, non-thermal radiation sources} 

\section{Introduction} \label{sec:intro}

High-energy astrophysical systems such as black holes and neutron stars often undergo transient events that release magnetized ejecta filled with hot collisionless plasma. Relativistic electrons and positrons in these ejecta are powerful emitters of synchrotron radiation. Understanding the evolution and emissive properties of such plasmas is important for modeling and interpreting astronomical observations.

Synchrotron emission is strongly influenced by the particle pitch-angle distribution, with the radiative power strongest from particles with large pitch angles (i.e., those with momenta predominantly perpendicular to the direction of the magnetic field). As a consequence, synchrotron cooling reduces the perpendicular component of the plasma pressure and thereby drives pressure anisotropy. If the plasma beta (thermal-to-magnetic pressure ratio) is sufficiently high, then the plasma will naturally become unstable to the firehose instability~\citep{chandrasekhar_1958,parker_1958,vedenov_1958}.  While the firehose instability has been studied extensively in the astrophysical context \citep[e.g.,][]{schekochihin_etal_2010,rosin_etal_2011,kunz_etal_2014, ley_etal_2022} and the solar wind \citep[e.g.,][]{hellinger_etal_2015}, its role in radiative, relativistic plasmas relevant to high-energy astrophysical systems remains to be explored. We argue that accounting self-consistently for the synchrotron radiative cooling process, and for the potentially unstable distribution function it drives, is essential for understanding the nature and radiative signatures of high-energy astrophysical plasmas.

In this work, we investigate (both analytically and numerically) the onset, linear growth, and nonlinear saturation of the firehose instability triggered by synchrotron cooling, to which we refer as the {\it synchrotron firehose instability} (SFHI). We show that, as an initially isotropic particle distribution undergoes synchrotron cooling (in an optically thin medium), it develops an anisotropic, nonthermal shape that becomes unstable to the firehose instability at the classically expected thresholds. We perform 2D particle-in-cell (PIC) simulations to confirm that the firehose instability occurs near the predicted thresholds; subsequently, it redistributes particles in pitch angle and thereby leads to a weakly anisotropic, near-thermal distribution. This mixing process brings energetic particles with small initial pitch angles to large pitch angles, allowing them to participate in cooling, and thereby enhances the overall cooling rate of the plasma. More dramatically, the redistribution of particles will have has a major effect on the radiative signatures of the plasma, as observed from various lines of sight and energy bands, by rapidly depleting the nonthermal and anisotropic components expected from stable cooling evolution. In PIC simulations, we study the nonlinear evolution of the SFHI in both relativistic electron-positron (pair) plasmas and electron-ion plasmas. We find that the SFHI is a viable collisionless mechanism of electron-ion thermal coupling when the ions are relativistic. 

The energetics of the SFHI in an electron-ion plasma can be summarized as follows. The initial plasma, taken to be isotropic and uniform, has no free energy and is at maximum entropy. However, synchrotron cooling reduces the electron kinetic energy and entropy by transferring them to photons (which escape the optically thin system), building up free energy in the electron pressure anisotropy ($P_{\parallel,e} > P_{\perp,e}$, where $P_{\parallel,e}$ and $P_{\perp,e}$ are pressure components parallel and perpendicular to the magnetic field, respectively). Once the associated electron pressure anisotropy becomes sufficiently large, it is tapped by firehose-unstable magnetic fields, which smooth the anisotropies in the electron distribution (thus increasing entropy) by scattering energetic electrons from low to high pitch angles. Interestingly, if the ions are relativistic (and thus have similar gyroradii as the electrons), these firehose fluctuations also scatter the ions from low pitch angles to high pitch angles, building up an ion pressure anisotropy (and thus free energy) in the opposite direction from the electrons (i.e, with $P_{\parallel,i} < P_{\perp,i}$). Once sufficient ion pressure anisotropy develops, the associated free energy is released into magnetic energy; this transfer can be mediated either by the ion mirror instability or ion-cyclotron instability, with our PIC simulations indicating the latter scenario. A net transfer of internal energy from ions to electrons is then enabled via dissipation of the ion-cyclotron magnetic fields, causing the ions to cool in tandem with electrons.

In Section~\ref{sec:theory}, we derive the analytical solution for the stable evolution of a synchrotron-cooling distribution of relativistic particles from an initially isotropic state, and describe analytical expectations for the onset and growth of the firehose instability based on linear theory. In Section~\ref{sec:num}, we employ PIC simulations to confirm the onset of the SFHI (close to the theoretical thresholds) and study the nonlinear evolution. In Section~\ref{sec:disc}, we discuss the astrophysical implications of our work. Finally, we conclude in Section~\ref{sec:conc}. Appendix~\ref{appendix} contains analytical calculations of the linear SFHI from the relativistic Vlasov--Maxwell equations.

\section{Analytical results} \label{sec:theory}

In this section, we first describe the analytical solution for the evolution of a population of synchrotron-cooling relativistic particles in a uniform magnetized plasma (\S\ref{sec:stable}). We then argue that this analytical solution eventually becomes unstable to the firehose instability if the plasma beta is sufficiently high (\S\ref{sec:crit}). If the plasma beta is low, however, we anticipate that the solution will remain valid indefinitely, until other physical effects intercede.

\subsection{Stable cooling evolution} \label{sec:stable}

We consider a population of electrons (or positrons) with ultra-relativistic temperature (such that momenta $p\gg m_e c$) being cooled by synchrotron radiation. The classical synchrotron radiation reaction force can be expressed as
\begin{equation}
\boldsymbol{F}_{\rm sync} = - \frac{2}{3} r_e^2 \gamma^2 \left[ \left( \boldsymbol{E} + \frac{\boldsymbol{v}}{c} \boldsymbol{\times B} \right)^2 - \left( \frac{\boldsymbol{v \cdot E}}{c} \right)^2 \right] \frac{\boldsymbol{v}}{c} , \label{eq:rr}
\end{equation}
where $r_e \equiv e^2/m_e c^2$ is the classical electron radius, $\boldsymbol{v} = \boldsymbol{p}/(\gamma m_e)$ is the particle velocity, and $\gamma \equiv (1-v^2/c^2)^{-1/2}$ is the Lorentz factor \citep{rybicki_lightman_1991}. Specializing to the case of a uniform magnetic field $\boldsymbol{B} = \boldsymbol{B}_0$ and vanishing electric field $\boldsymbol{E} = 0$, Eq.~\eqref{eq:rr} can be conveniently written as
\begin{equation}
\boldsymbol{F}_{\rm sync} = 
- \frac{2}{3} \frac{r_e^2 B_0^2}{m_e^2 c^2} p^2 \sin^2{\theta} \,\hat{\boldsymbol{p}} = 
-\, \frac{p^2 \sin^2{\theta}}{\overline{p}_0 \tau_{\rm cool}} \,\hat{\boldsymbol{p}} , \label{eq:rrred}
\end{equation}
where $\theta \equiv \cos^{-1}(\boldsymbol{B \cdot p}/ B p)$ is the particle's pitch angle, $\overline{p}_0$ is the initial average momentum, and
\begin{equation}
    \tau_{\rm cool} \equiv \frac{3 m_e^2 c^2}{2 r_e^2 B_0^2 \overline{p}_0}
\end{equation}
is the characteristic cooling timescale for the average particle. 

\begin{figure}
   \includegraphics[width=\columnwidth]{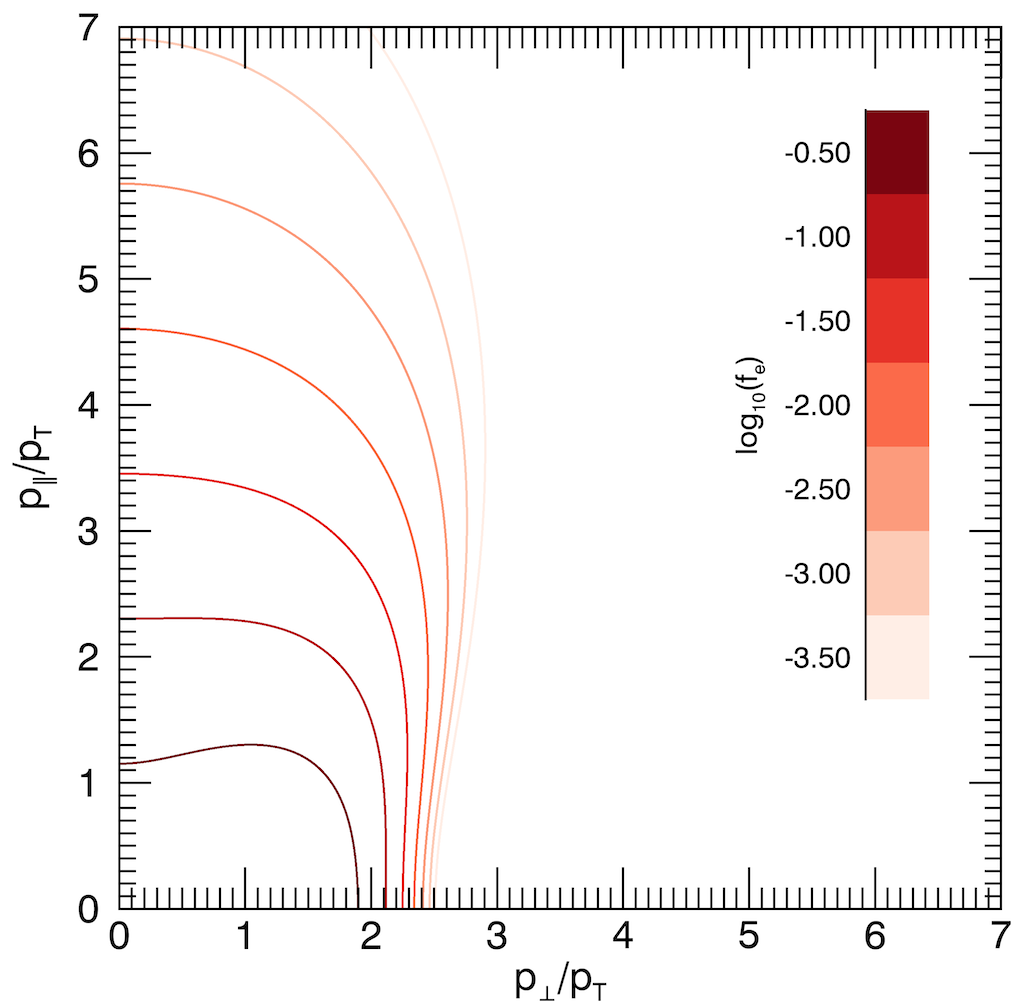}
   \includegraphics[width=\columnwidth]{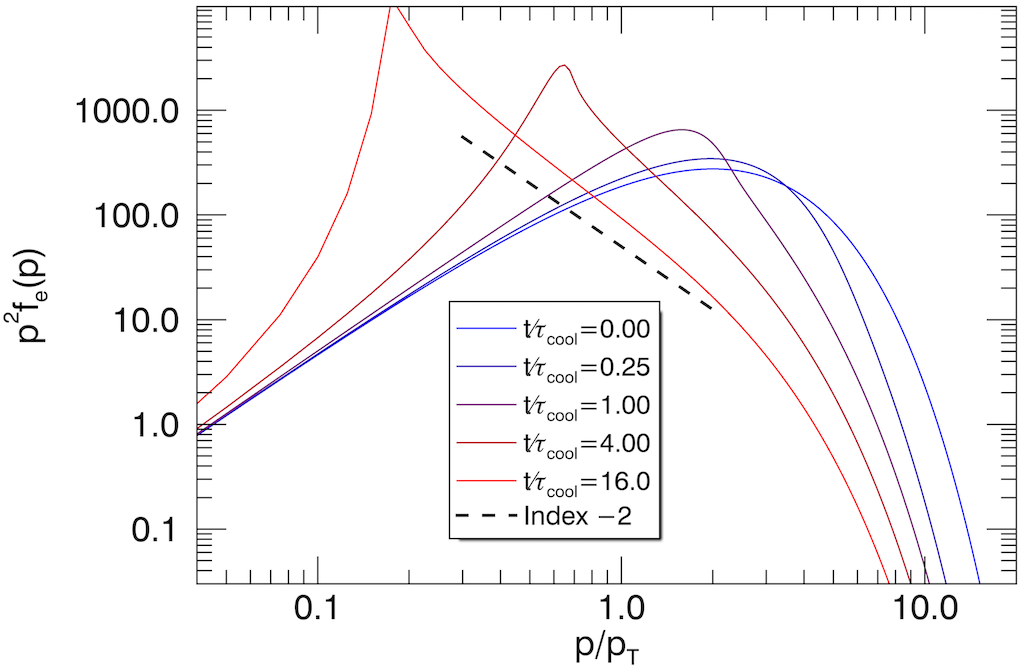}
   \includegraphics[width=\columnwidth]{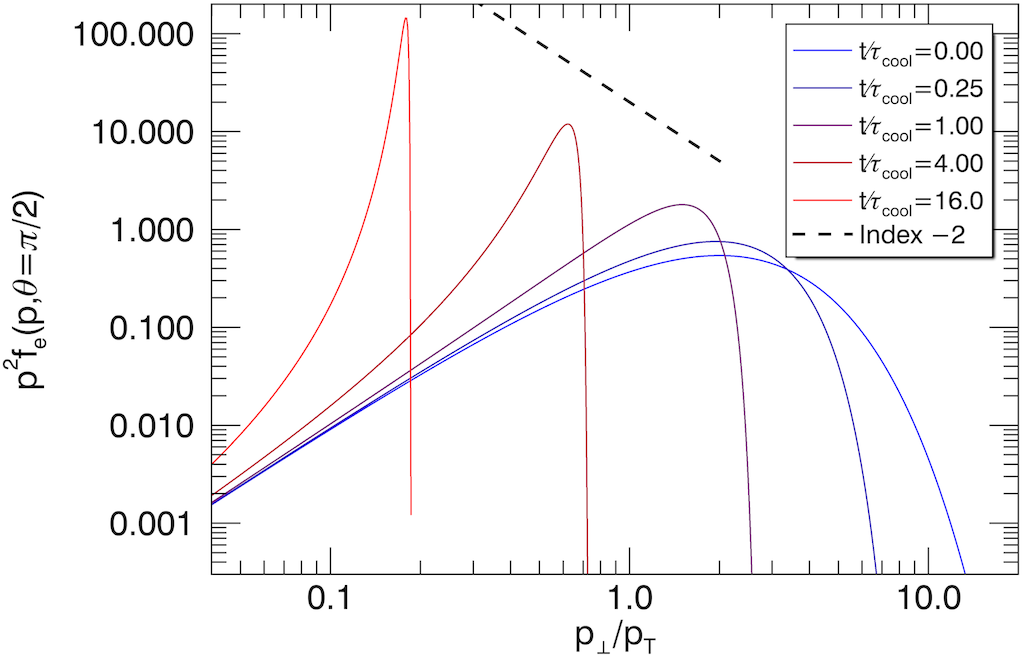}
   \centering
   \caption{\label{fig:analytic} Top panel: Contours of the synchrotron-cooling electron distribution $f_e(p,\theta)$ from the analytic solution \eqref{eq:sol} at $t = \tau_{\rm cool}$. Middle panel: evolution of the angle-integrated momentum distribution $p^2 f_e(p)$; a power-law with index $-2$ is shown for reference. Bottom panel: the distribution $p^2 f_e(p,\theta)$ along the perpendicular direction ($\theta = \pi/2$) for the same times.}
 \end{figure}

The evolution of a uniform, gyrotropic electron distribution $f_e(\boldsymbol{p},t) = f_e(p,\theta,t)$ subject to the force \eqref{eq:rrred} is described by the Vlasov equation,
\begin{equation}
\partial_t f_e = - \frac{\partial}{\partial \boldsymbol{p}} {\bcdot} \left( \boldsymbol{F}_{\rm sync} f_e \right) = \frac{\sin^2{\theta}}{\tau_{\rm cool} \overline{p}_0 p^2} \,\partial_p \left( p^4 f_e \right) .
\end{equation}
Given an isotropic initial distribution $f_{e0}(p)$, this equation can be readily solved by the method of characteristics to obtain
\begin{equation}
f_e(p,\theta,t) = \frac{f_{e0}\bigl(p/[1 - p t \sin^2{\theta}/(\overline{p}_0 \tau_{\rm cool})]\bigr)}{[1- p t \sin^2{\theta}/(\overline{p}_0 \tau_{\rm cool})]^4 } \label{eq:gensol}
\end{equation}
for $p < p_{\rm cut}(\theta) \equiv \overline{p}_0 \tau_{\rm cool}/(t \sin^2{\theta})$ and $f_e = 0$ for $p \ge p_{\rm cut}(\theta)$. For an initial ultra-relativistic Maxwell--J\"{u}ttner distribution,~$f_{e0} = n_0 \exp{(-p/p_T)}/(8\pi p_T^3)$, Eq.~\eqref{eq:gensol} becomes
\begin{equation}
f_e(p,\theta,t) = \frac{n_0}{8\pi p_T^3} \frac{\exp{\biggl[ \dfrac{-p/p_T}{1 - p t \sin^2{\theta}/(3 p_T \tau_{\rm cool})} \biggr]}}{\bigl[1 - p t \sin^2{\theta}/(3 p_T \tau_{\rm cool})\bigr]^4} , \label{eq:sol}
\end{equation}
where $p_T \equiv T_{e0}/c = \overline{p}_0/3$ is the thermal momentum and $n_0$ is the number density. The solutions given by Eqs.~\eqref{eq:gensol}--\eqref{eq:sol} have a hard cutoff at $p = p_{\rm cut}(\theta)$, due to high-energy particles catching up to low-energy particles as they cool (causing a ``phase-space shock''). The maximum of the distribution \eqref{eq:sol} at a given momentum can be shown (from $\partial f_e / \partial \theta = 0$) to occur at a pitch angle satisfying $\sin^2{\theta} = 3 (p_T/p - 1/4 ) (\tau_{\rm cool}/t)$ for $(1/4 + t/3\tau_{\rm cool})^{-1} < p/p_T < 4$, at $\theta=0$ for $p/p_T > 4$, and at $\theta=\pi/2$ for $p/p_T < (1/4 + t/3\tau_{\rm cool})^{-1}$; thus high-energy particles are focused in the direction of~$\boldsymbol{B}_0$. Iso-contours of the solution given by Eq.~\eqref{eq:sol} at $t = \tau_{\rm cool}$ are plotted in the top panel of Fig.~\ref{fig:analytic}. In the middle panel of Fig.~\ref{fig:analytic}, we show the distribution averaged over angles, $p^2 f_e(p,t) \equiv (1/2) p^2 \int\rmd\cos{\theta}\, f_e(p,\theta,t)$, at times $t/\tau_{\rm cool} \in \{ 0, 0.25, 1, 4, 16 \}$. At late times, the solution develops a power-law range with scaling $p^2 f_e(p) \propto p^{-2}$ from $p \sim p_{\rm cut}(\pi/2)$ to $p \sim \overline{p}_0$, which is a consequence of the superposition of phase-space shocks at different $p_{\rm cut}(\theta)$. In the bottom panel of Fig.~\ref{fig:analytic}, we show the evolution of the distribution in the perpendicular direction, $p^2 f_e(p,\theta=\pi/2)$, highlighting the gradual steepening of the phase-space shock.

For early times $\epsilon \equiv t/\tau_{\rm cool} \ll 1$, Eq.~\eqref{eq:sol} can be Taylor-expanded as long as the momenta considered are not too large, $p \lesssim p_T$. To first order in $\epsilon$, the result is
\begin{equation}
f_e \approx \frac{n_0}{8 \pi p_T^3} \left[ 1 + \left( 4 - \frac{p}{p_T} \right) \frac{p t}{3 p_T \tau_{\rm cool}} \sin^2{\theta} \right] {\rm e}^{- p/p_T} \,. \label{eq:solexp}
\end{equation}
The distribution retains its shape in the parallel direction ($\theta=0$), while in the perpendicular direction ($\theta=\pi/2$) it has a pile-up of particles for $p < 4 p_T$ and a depletion of particles for $p > 4 p_T$.

The evolution of the electron kinetic energy density $n_0 E_{{\rm kin},e}$ and the parallel and perpendicular components of the electron pressure tensor may be obtained by taking the appropriate moments of Eq.~\eqref{eq:solexp}. The results are
\begin{align}
n_0 E_{{\rm kin},e} &= \int\rmd^3 p \, p c f_e \approx  3 n_0 p_T c \left(1 - \frac{8}{9} \frac{t}{\tau_{\rm cool}} \right) \,,\nonumber \\ 
P_{\parallel,e} &= \int \rmd^3p \, p_\parallel v_\parallel f_e \approx n_0 p_T c \left( 1 - \frac{8}{15} \frac{t}{\tau_{\rm cool}} \right) \,, \nonumber \\ 
P_{\perp,e} &= \frac{1}{2} \int \rmd^3p \, p_\perp v_\perp f_e \approx  n_0 p_T c \left( 1 - \frac{16}{15}  \frac{t}{\tau_{\rm cool}} \right) . \label{eq:pressures} 
\end{align}
where the ``$\parallel$'' and ``$\perp$'' subscripts indicate directions parallel and perpendicular to $\boldsymbol{B}_0$. As a consequence, to first order in $\epsilon$, the pressure anisotropy is $P_{\perp,e}/P_{\parallel,e} \approx 1 - (8/15) (t/\tau_{\rm cool})$.

 In Sec.~\ref{sec:num}, we reproduce numerically the analytical solution of Eq.~\eqref{eq:sol} by performing a PIC simulation of uncharged, synchrotron-cooling gas that remains stable to electromagnetic instabilities.

\subsection{Firehose instability criterion} \label{sec:crit}
 
The solution~\eqref{eq:sol} will be valid until the onset of the firehose instability, which occurs once sufficient pressure anisotropy is produced to offset the magnetic tension. We anticipate that the instability will grow initially at a super-exponential rate, since the firehose growth rate $\gamma_{\rm f} \equiv \rmd\ln{B_{\rm f}}/\rmd t$ will increase gradually from $\gamma_{\rm f}=0$ at marginal stability to a value $\gamma_{\rm f} \gtrsim \tau_{\rm cool}^{-1}$ as the plasma evolves into the unstable region of parameter space via continuous cooling and corresponding growth of the pressure anisotropy (unregulated by the instability). Here, we define $B_{\rm f}(t)$ as the amplitude of the most unstable firehose mode. This super-exponential (or ``unregulated'') stage may be followed by an exponential (or ``regulated'') stage once the linear growth rate $\gamma_{\rm f}$ becomes large enough that the firehose fluctuations deplete the pressure anisotropy faster than it is replenished by the synchrotron cooling. In practice, the regulated stage may be relatively brief compared to the unregulated stage.
 
In Appendix~\ref{appendix}, we calculate analytically the parallel firehose instability criterion from the linearized relativistic Vlasov--Maxwell equations in the limit of long wavelength and low frequency (relative to kinetic scales), using the early-time ($t/\tau_{\rm cool} \ll 1$) distribution from Eq.~\eqref{eq:solexp} as a (slowly-varying) background. This instability criterion is
\begin{align}
\frac{P_{\perp,e}}{P_{\parallel,e}} < 1 - \frac{C_{\rm thr}}{\beta_{\parallel,e}} , \label{eq:fhcrit}
\end{align}
where $C_{\rm thr}$ is an order-unity constant and $\beta_{\parallel,e}\equiv 8\pi P_{\parallel,e}/B^2$ is the plasma beta using the parallel component of the electron pressure. In our calculation, $C_{\rm thr} = 1$ for a relativistic pair plasma (where positrons have the same background cooling distribution as electrons) and $C_{\rm thr}=2$ for an electron-ion plasma (where ions do not cool and thus retain their initial Maxwell--J\"{u}ttner distribution $f_{i0}$ as the background). Eq.~\eqref{eq:fhcrit} agrees with the standard criterion for the relativistic fluid firehose instability \citep[e.g.,][]{barnes_scargle_1973}. The inequality \eqref{eq:fhcrit} is satisfied for times after an ``onset'' time,
\begin{align}
    t \gtrsim \tau_{\rm onset} \equiv \frac{15 C_{\rm thr}}{8\beta_{e0}} \tau_{\rm cool} ,
\end{align}
based on the early-time expressions in Eq.~\eqref{eq:pressures}; this is valid for $\beta_{e0} \gg 1$, where $\beta_{e0}$ is the initial electron plasma beta. In terms of physical parameters, the onset time can be expressed as
\begin{equation}
    \tau_{\rm onset}= \frac{5C_{\rm thr}}{16} \bigl(\theta_{e0}^2 \sigma_{\rm T} n_0 c\bigr)^{-1}, \label{eq:tauonset}    
\end{equation}
where $\theta_{e0} = T_{e0}/m_e c^2$ is the dimensionless temperature and $\sigma_{\rm T} = (8\pi/3) r_e^2$ is the Thomson cross-section. Notably, $\tau_{\rm onset}$ is independent of the magnetic-field strength~$B_0$. For comparison, the (relativistic) Coulomb collision time also scales in proportion to $(\sigma_{\rm T} n_0 c)^{-1}$, but with a factor $\theta_{e0}^2$ rather than $\theta_{e0}^{-2}$ \citep{stepney_1983}, so that the onset time is a factor of $\theta_{e0}^{-4}$ shorter than the collision time. Thus, we expect Coulomb collisions to be negligible over the onset timescale, as long as electrons are sufficiently relativistic.

Based on previous studies of the firehose instability in the non-relativistic regime, we anticipate that finite-Larmor-radius corrections will reduce $C_{\rm thr}$ by a factor ${\approx}0.7$, making the firehose easier to trigger \citep[e.g.,][]{hellinger_matsumoto_2000, bott_etal_2021}. We also anticipate that, at least in some physical regimes, the firehose instability may occur at oblique rather than parallel orientation \citep[e.g.,][]{li_habbal_2000}.

\begin{figure}
   \includegraphics[width=\columnwidth]{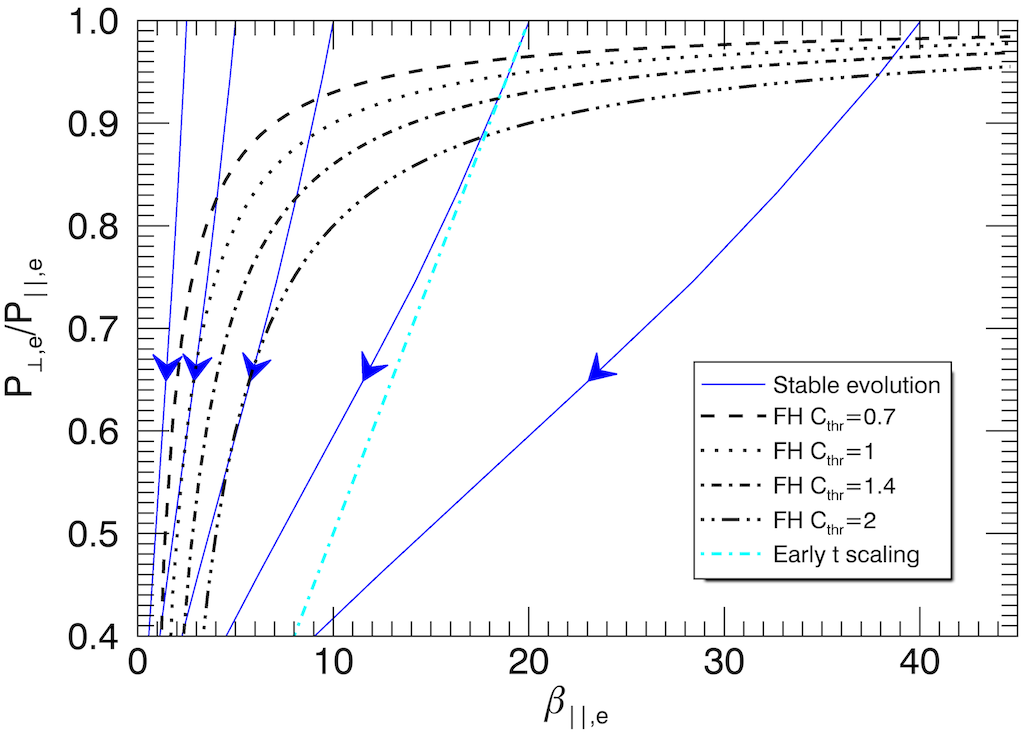}
   \centering
   \caption{\label{fig:brazil_analytic} Evolution of the analytical solution \eqref{eq:sol} for the synchrotron-cooling electron distribution, in the $P_{\perp,e}/P_{\parallel,e}$ versus $\beta_{\parallel,e}$ parameter space; trajectories for several values of initial beta are shown, $\beta_{\parallel,e,0} \in \{2.5, 5, 10, 20, 40 \}$ (blue). Arrowheads indicate the direction of evolution. For comparison, the early-time approximation \eqref{eq:pressures} is shown for $\beta_{\parallel,e,0}=20$ (cyan). The firehose instability thresholds \eqref{eq:fhcrit} for $C_{\rm thr} \in \{0.7, 1, 1.4, 2\}$ are also shown (black; see legend).}
 \end{figure}
 
 To illustrate the relevant region of the firehose instability in parameter space, we employ plots of $P_{\perp,e}/P_{\parallel,e}$ versus $\beta_{\parallel,e}$. In Fig.~\ref{fig:brazil_analytic}, we show the trajectory of the synchrotron cooling distribution [obtained by integrating Eq.~\eqref{eq:sol} numerically] across
 this parameter space, for initial values of $\beta_{\parallel,e,0} \in \{2.5, 5, 10, 20, 40 \}$ at isotropy. The displayed solutions cover a duration of $t/\tau_{\rm cool} \approx 16$, at which time $P_{\perp,e}/P_{\parallel,e} \approx 0.4$. We also show the firehose instability thresholds from Eq.~\eqref{eq:fhcrit} with $C_{\rm thr} \in \{0.7, 1, 1.4, 2\}$. The $\beta_{\parallel,e,0}=2.5$ case misses all of the instability thresholds, while $\beta_{\parallel,e,0}>10$ is sufficient to hit all of the thresholds (after which the development of the SFHI will invalidate the analytical solution). This motivates choosing $\beta_{\parallel,0} = 20$ as the fiducial value in the PIC simulations of Sec.~\ref{sec:num}, which allows a robust incursion into the unstable zone while avoiding extreme values of $\beta$ (which are unrealistic and difficult to simulate with PIC).
 
\subsection{Firehose growth} \label{sec:growth}
 
Once the SFHI growth rate becomes positive (at $t = \tau_{\rm onset}$), we anticipate that the (fast) growth of the most unstable firehose mode at any instant can be modeled using linear theory, but that this growth rate will be modulated on the (slow) cooling timescale due to the continual growth of pressure anisotropy (until regulation). The linear calculation in Appendix~\ref{appendix} suggests a growth rate of the form
\begin{equation}
    \gamma_{\rm f} \equiv \frac{\rmd\ln{B}_{\rm f}}{{\rm d} t} = \zeta k c \left( \frac{\delta t}{\tau_{\rm cool}} \right)^{1/2} , \label{eq:gamma_theory}
\end{equation}
where $\zeta$ is a numerical coefficient that depends on the plasma composition and temperature [see Eqs.~\eqref{eq:gammaurei}, \eqref{eq:gammasrei}, and \eqref{eq:gammapair} in Appendix~\ref{appendix}] and $\delta t = t - \tau_{\rm onset}$ is the time after onset. We can integrate Eq.~\eqref{eq:gamma_theory} in time to obtain a super-exponential growth,
\begin{equation}
    B_{\rm f} = B_{\rm seed} \exp{\left[ \left( \frac{\delta t}{\tau_{\rm gr}} \right)^{3/2} \right]} , \label{eq:superexp}
\end{equation}
where $\tau_{\rm gr} = (2\zeta/3)^{-2/3} \tau_{\rm cool}^{1/3} (kc)^{-2/3}$ and $B_{\rm seed}$ is the initial seed perturbation magnitude \citep[see, e.g.,][for similar arguments applied to the spontaneous development of Weibel instability]{zhou_etal_2022}. Based on the literature, we expect the most unstable mode to be near the electron kinetic scales; specifically, $k \sim \rho_{e0}^{-1}$ for the electron firehose instability \citep[e.g.,][]{li_habbal_2000}, where $\rho_{e0}=3T_{e0}/eB_0$ is the characteristic ultra-relativistic electron Larmor radius. Therefore, ignoring coefficients of order unity, the unregulated growth occurs on a hybrid timescale $\tau_{\rm gr} \sim \tau_{\rm cool}^{1/3} \tau_{\rm gyro}^{2/3}$, where $\tau_{\rm gyro} = \rho_{e0}/c$ is the electron Larmor timescale. Because the unregulated growth is super-exponential, it is also on this timescale that the pressure anisotropy will be regulated close to the firehose threshold.

 \section{Numerical results} \label{sec:num}
 
 \subsection{Numerical setup}
 
To demonstrate the existence of the SFHI and study its properties, we perform a set of radiative PIC simulations using the code {\em Zeltron} \citep{cerutti_etal_2013}. The simulations include the synchrotron radiation reaction force, Eq.~\eqref{eq:rr}, in the particle pusher for electrons and positrons.

Because two spatial dimensions are sufficient to capture the basic features of the SFHI (with either parallel or oblique polarization), we limit ourselves to 2D simulations (an $x$--$y$ plane with mean magnetic field $\boldsymbol{B}_0 = B_0 \hat{\boldsymbol{y}}$ oriented in the domain), which allows us to run simulations with sufficiently long duration to study the slow-cooling regime ($\tau_{\rm gyro}/\tau_{\rm cool} \ll 1$). The spatial domain has periodic boundary conditions.

We initialize the PIC simulations with a uniform, isotropic thermal plasma of electrons and ions (or positrons). In all cases, we initialize both species with the same temperature, $T_{i0}/T_{e0} = 1$ (where the subscript $i$ refers to either ions or positrons). To study the effect of the plasma composition on the instability, we consider four cases: 
(1) an uncharged gas of ultra-relativistic particles, which undergo synchrotron cooling but no electromagnetic interactions, so the instability is artificially inhibited and thus the distribution reproduces the stable analytical solution described in Sec.~\ref{sec:stable}; (2) an ultra-relativistic pair plasma ($m_i/m_e = 1$) with dimensionless temperature $\theta_{e0} \equiv T_{e0}/m_e c^2 = 100$, where both species radiate; (3) an electron-ion plasma (with ``ions" meaning protons, $m_i/m_e = 1836$) with sub-relativistic ions, taking dimensionless ion temperature $\theta_{i0} \equiv T_{i0}/m_i c^2 = 1/64 \ll 1$, such that electrons are in the ultra-relativistic regime with $\theta_{e0} = (m_i/m_e) \theta_{i0} \approx 29 \gg 1$ (we adopt the terminology in \citealt{werner_etal_2018} to call this the ``semirelativistic" regime);  and (4) an ultra-relativistic electron-ion plasma ($m_i/m_e = 1836$) with $\theta_{i0} = 100$ (and thus $\theta_{e0} \gg 1$ as well).

Our fiducial parameters are $\beta_0 \equiv 8 \pi n_0 (T_{i0}+T_{e0})/B_0^2 = 40$ [such that $\tau_{\rm onset}/\tau_{\rm cool} = 3C_{\rm thr}/32$ from Eq.~\eqref{eq:fhcrit}] and $\tau_{\rm gyro}/\tau_{\rm cool} = 1.2\times 10^{-4}$. The fiducial domain size is set to $L/2\pi \rho_{e0} \approx 4$, sufficient to contain the most unstable electron firehose mode; however, we also performed shorter-duration simulations with $L/2\pi\rho_{e0} \approx 8$ that give similar results and are used for some of the analysis. We also ran the nonrelativistic ion case on a domain that is four times larger ($L/2\pi\rho_{e0} \approx 16$, $L/2\pi\rho_{i0} \approx 3.5$ where the non-relativistic ion gyroradius is $\rho_{i0} \approx \sqrt{m_i T_{i0}}c/eB_0 \approx 4.7 \rho_{e0}$ in this case) with shorter duration, to confirm that ion-scale dynamics do not influence the results. The fiducial simulation lattice is $512^2$ cells (with cell size $\Delta x \approx \rho_{e0}/20$) and the number of particles per cell per species is~$512$; the larger cases have the same resolution but correspondingly more cells. The simulations have a timestep $\Delta t \approx \Delta x/(\sqrt{3}c)$ and fiducial duration of $t/\tau_{\rm cool} \sim 3$.

\subsection{Electron firehose dynamics} \label{sec:electrons}

  \begin{figure}
     \includegraphics[width=\columnwidth]{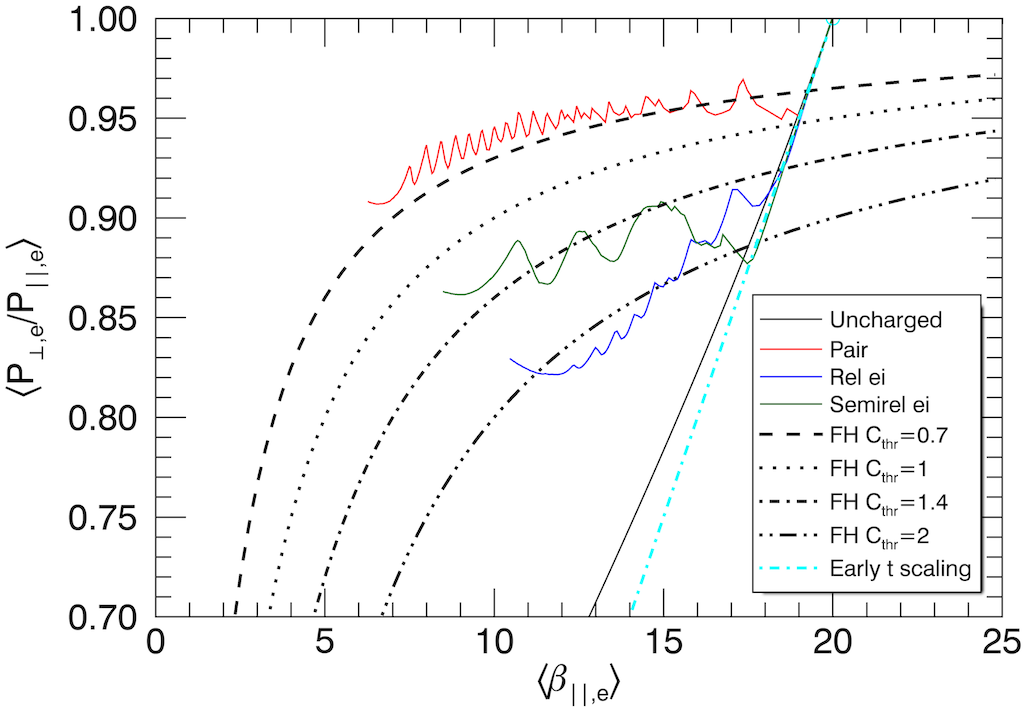}
   \includegraphics[width=\columnwidth]{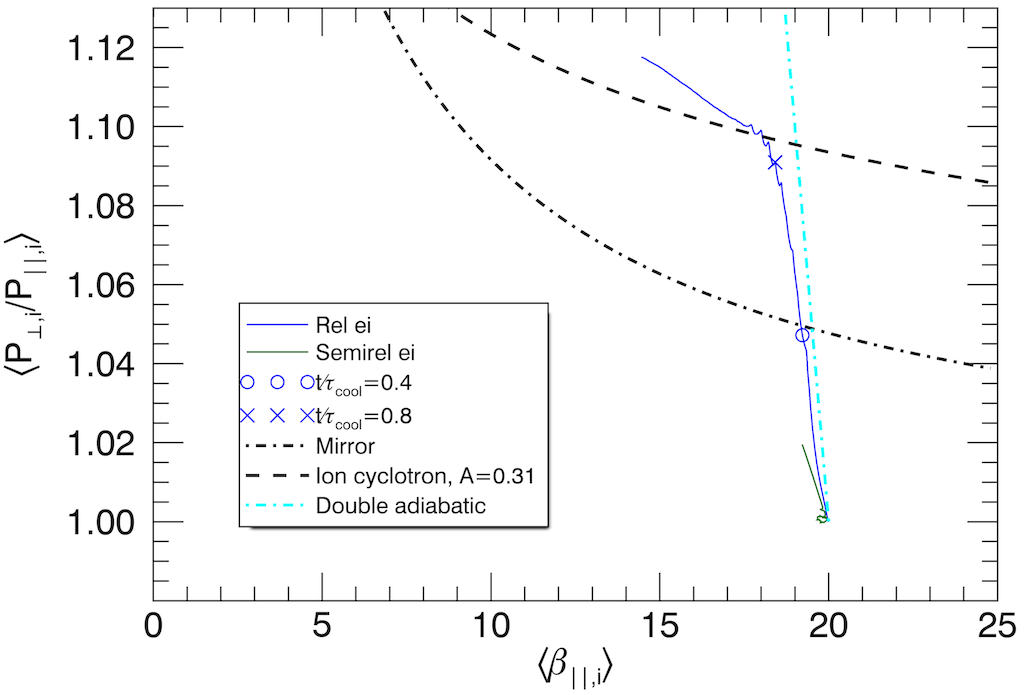}
   \centering
   \caption{\label{fig:brazil} Top panel: average electron pressure anisotropy $\langle P_{\perp,e}/P_{\parallel,e} \rangle$ versus $\langle \beta_{\parallel,e}\rangle$ for different plasma compositions: uncharged plasma (black), ultra-relativistic pair plasma (red), semirelativistic electron-ion plasma (green), and ultra-relativistic electron-ion plasma (blue). Instability thresholds for several values of $C_{\rm thr}$ are shown by the dashed/dotted lines, as indicated in the legend. The early-time analytic solution for stable evolution from Eq.~\eqref{eq:pressures} is also shown (cyan). Bottom panel: similar for ions in the electron-ion simulations, $\langle P_{\perp,i}/P_{\parallel,i} \rangle$ versus $\langle \beta_{\parallel,i} \rangle$. Times of $t/\tau_{\rm cool} = 0.4$ (circle) and $t/\tau_{\rm cool} = 0.8$ (cross) are marked on the trajectory for the relativistic ion case. Double adiabatic evolution [cyan, Eq.~\eqref{eq:adi_scaling}] is also shown for reference, along with mirror and ion-cyclotron instability thresholds described in the text (black lines).}
 \end{figure}
 
In all cases, the plasma initially undergoes a stable cooling phase, in which both the parallel and perpendicular components of electron (and/or positron) pressure decline, with the perpendicular one declining faster. When the pressure becomes sufficiently anisotropic, with $P_{\parallel,e}>P_{\perp,e}$, the plasma goes unstable to the SFHI (we confirmed separately that the instability is absent if the initial $\beta$ is too small, $\beta_0 \lesssim 1$). The evolution of the volume-averaged electron pressure ratio $\langle P_{\perp,e}/P_{\parallel,e} \rangle$ versus the volume-averaged electron plasma beta $\langle \beta_{e,\parallel} \rangle$ is shown in the top panel of Fig.~\ref{fig:brazil} for all four compositions. In the three charged cases, the evolution of $\langle P_{\perp,e}/P_{\parallel,e} \rangle$ traces the expected early-time scaling $P_{\perp,e}/P_{\parallel,e} \sim 1 - (8/15)(t/\tau_{\rm cool})$ [Eq.\eqref{eq:pressures}] until the onset of the~SFHI. Thereafter, the decrease of $\langle P_{\perp,e}/P_{\parallel,e} \rangle$ stalls while $\beta_{e,\parallel}$ continues to decline, and the trajectory fluctuates along the corresponding instability threshold. The pair-plasma case lies near the instability threshold of Eq.~\eqref{eq:fhcrit} with $C_{\rm thr} \approx 0.7$ and the semi-relativistic electron-ion case near $C_{\rm thr} \approx 1.4$, in agreement with the analytical expectations from Sec.~\ref{sec:crit}. Meanwhile, the relativistic electron-ion case is closest to the threshold with $C_{\rm thr} \approx 2$, somewhat larger than the expectation of $C_{\rm thr} \approx 1.4$; one reason for this discrepancy may be the influence of thermal coupling between the electrons and relativistic ions, which is discussed in Sec.~\ref{sec:ions}.
 
   \begin{figure}
   \includegraphics[width=0.49\columnwidth]{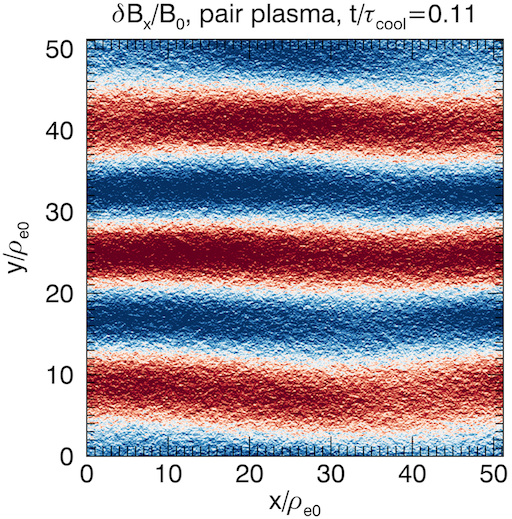}
   \includegraphics[width=0.49\columnwidth]{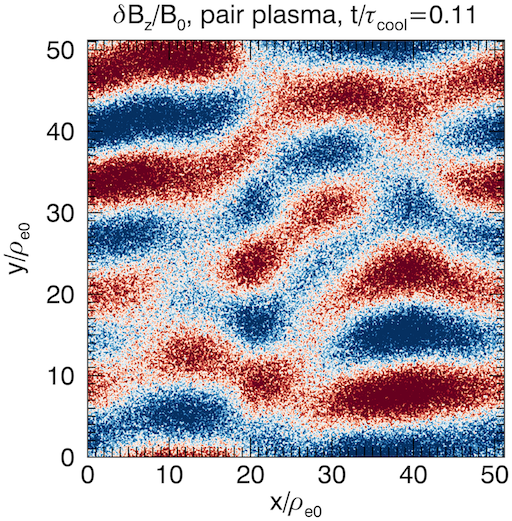} \\
   \includegraphics[width=0.49\columnwidth]{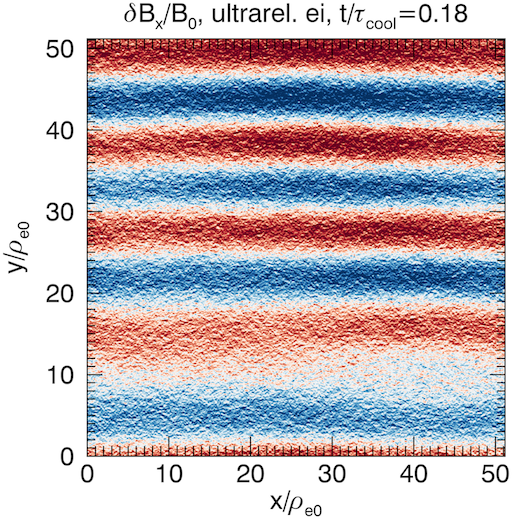}
   \includegraphics[width=0.49\columnwidth]{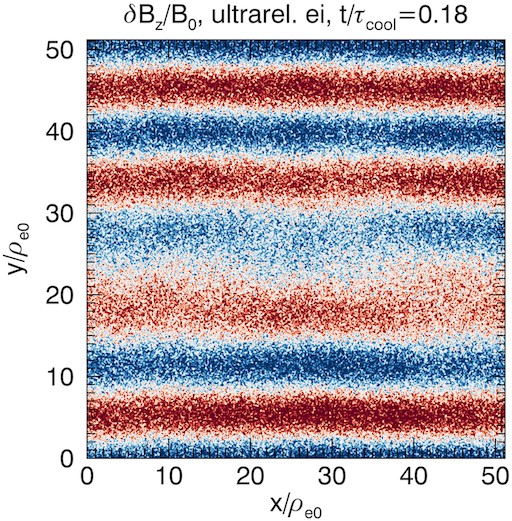} \\
   \includegraphics[width=0.49\columnwidth]{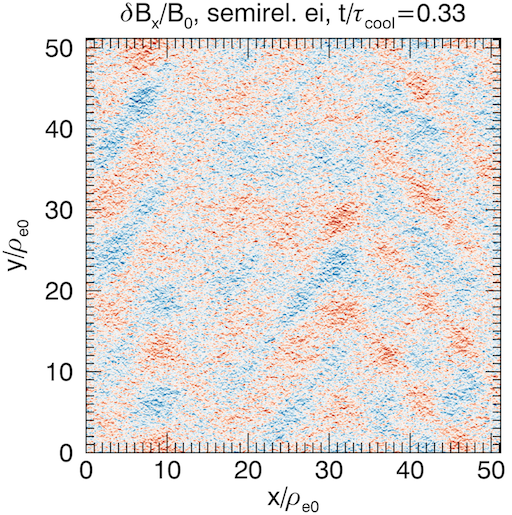}
   \includegraphics[width=0.49\columnwidth]{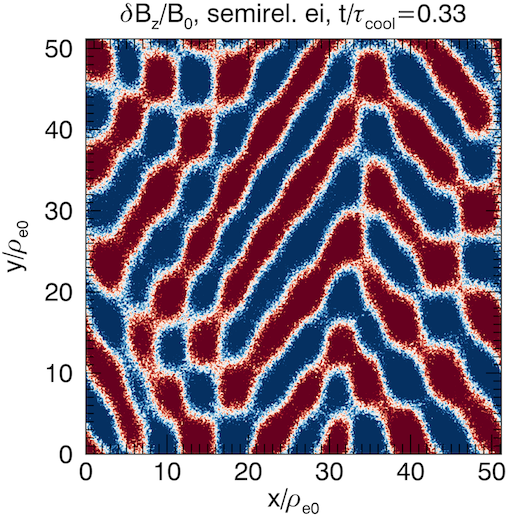} \\
   \centering
   \caption{\label{fig:prof} Images of perpendicular magnetic field fluctuations, $B_x$ (left column) and $B_z$ (right column), for pair plasma (top row), relativistic electron-ion plasma (middle row), and semirelativistic electron-ion plasma (bottom row). The snapshots are taken at times during the first cycle, indicated in the titles. The colormaps saturate at $\delta B/B_0 = 0.1$.}
 \end{figure}

In general, the firehose instability can be either parallel or oblique, with the dominant orientation determined by the physical parameters and plasma composition. In our PIC simulations, we find that the properties of the magnetic-field fluctuations produced by the SFHI depend on the plasma composition. The profiles of the perpendicular components $B_x$ and $B_z$ are shown in Fig.~\ref{fig:prof} for the three unstable compositions at double the fiducial size, taken at representative times after the onset of the instability (fluctuations in $B_y$ are negligible). The pair-plasma case (top row) has a mixture of parallel and oblique modes; the parallel mode is visible in the $B_x$ profile while the oblique mode distorts the $B_z$ profile. The relativistic electron-ion case (middle row) is dominated by circularly polarized parallel fluctuations, with no visible signatures of oblique modes. The case with sub-relativistic ions (bottom row), on the other hand, is strongly dominated by oblique modes (in~$B_z$), with no discernible sign of parallel modes (in~$B_x$); this is consistent with expectations from the non-relativistic regime, in which the oblique firehose generally grows faster than the parallel firehose \citep[e.g.,][]{li_habbal_2000, bott_etal_2021}.
  
   \begin{figure}
   \includegraphics[width=\columnwidth]{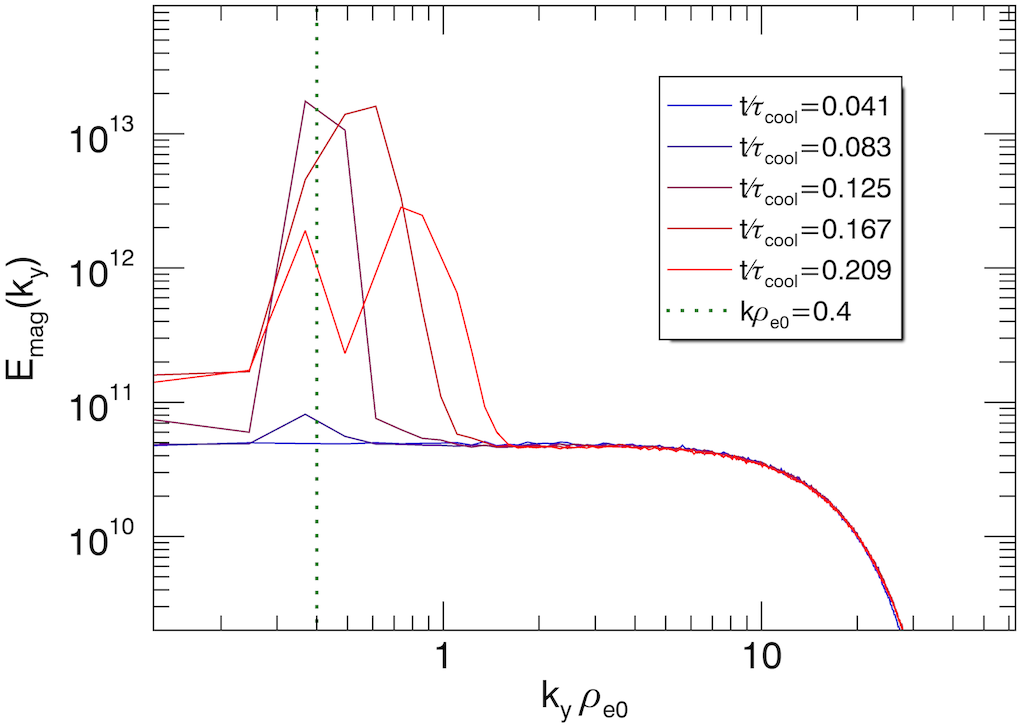} 
   \centering
   \caption{\label{fig:spec_mag} Magnetic energy spectrum for the pair-plasma case, averaged over time intervals between the times shown in the legend.}
 \end{figure}
 
The properties of the fluctuations also change over time: they typically start at a wavelength that is approximately twice larger than~$2\pi\rho_{e0}$, then distort and cascade to smaller-scale waves. In Fig.~\ref{fig:spec_mag}, we show the evolution of the magnetic energy spectrum (in the $k_y$ direction, integrated over $k_x$ and averaged over 5 time snapshots) at early times for the large pair-plasma case ($L/2\pi\rho_{e0} \approx 8$). The initial state (a flat spectrum) is due to numerical PIC noise. The early-time peak, during the linear stage, is at a wavenumber $k_y \approx 0.4/\rho_{e0}$, in agreement with previous works of the kinetic firehose instability \citep[e.g.,][]{kunz_etal_2014}. Note that the linear theory from Appendix~\ref{appendix} cannot be used to predict the most unstable wavenumber, due to the long-wavelength/low-frequency approximation. This then moves to higher wavenumbers and decays (the longer-term evolution of the system is described further below). There does not appear to be significant inverse transfer of energy to larger scales. The evolution of the spectrum is qualitatively similar for all of the electron-ion cases, peaking at a similar value of~$k\rho_{e0}$, which confirms the instability as being the electron firehose. In principle, since the electron pressure anisotropy causes an anisotropy in the total plasma pressure, it may trigger the fluid firehose instability, which would occur at ion scales $k_y \lesssim 1/\rho_i$ rather than the electron scales. However, this is not observed in our PIC simulations (even in the large-domain case having $L/2\pi\rho_{i0} \approx 3.5$), presumably because the electron firehose instability onset and growth is more rapid, and therefore depletes the pressure anisotropy before the ion-scale fluid firehose modes can grow.
 
  \begin{figure}
   \includegraphics[width=\columnwidth]{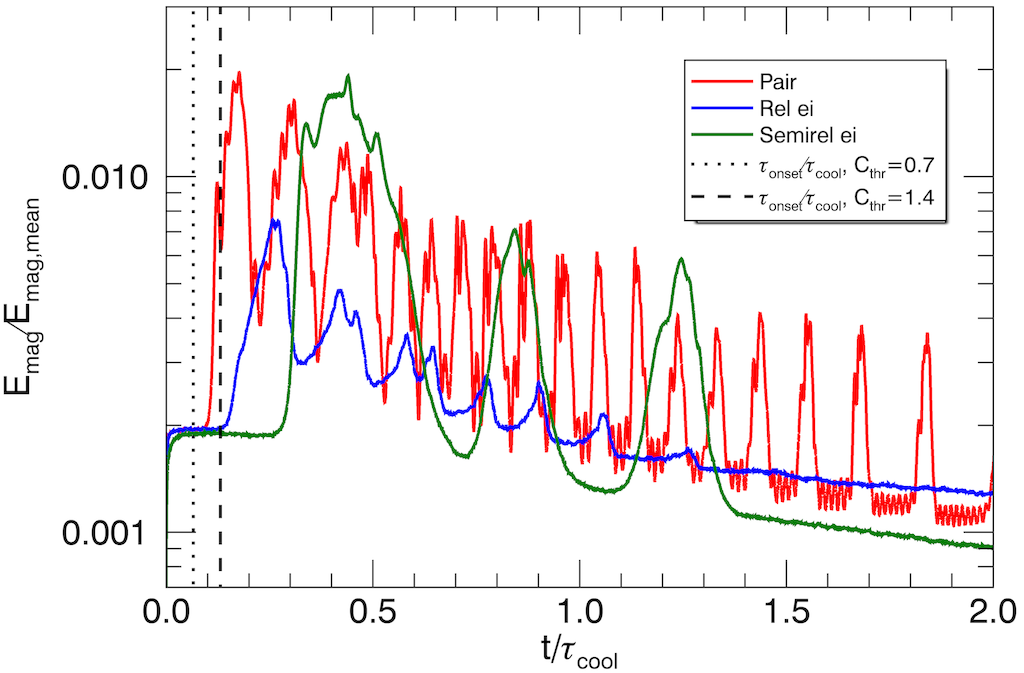} 
   \includegraphics[width=\columnwidth]{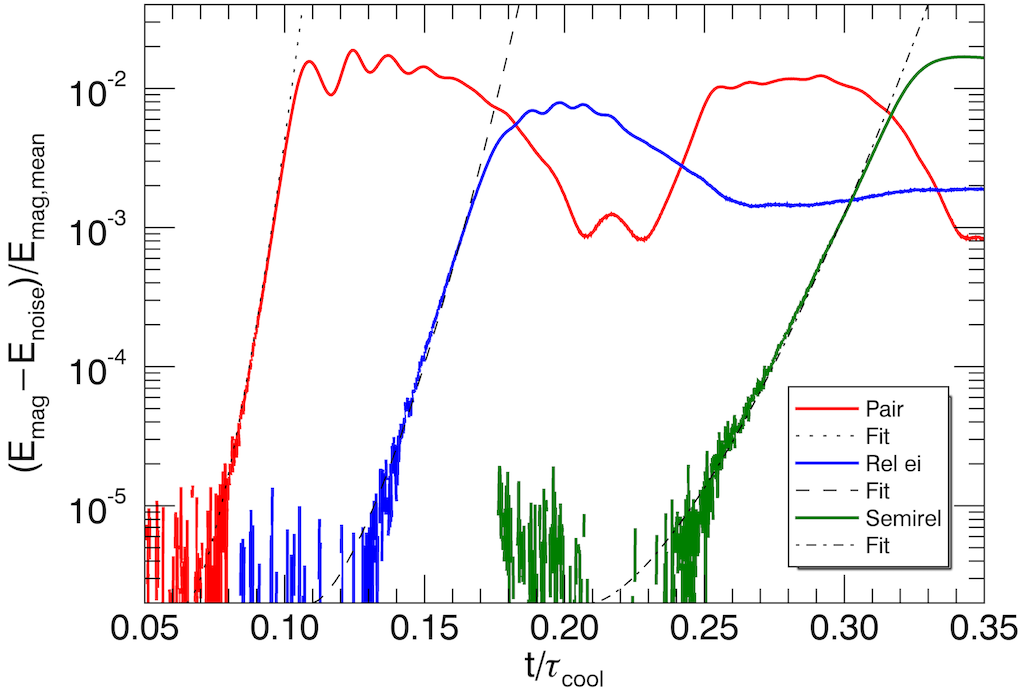}
  \includegraphics[width=\columnwidth]{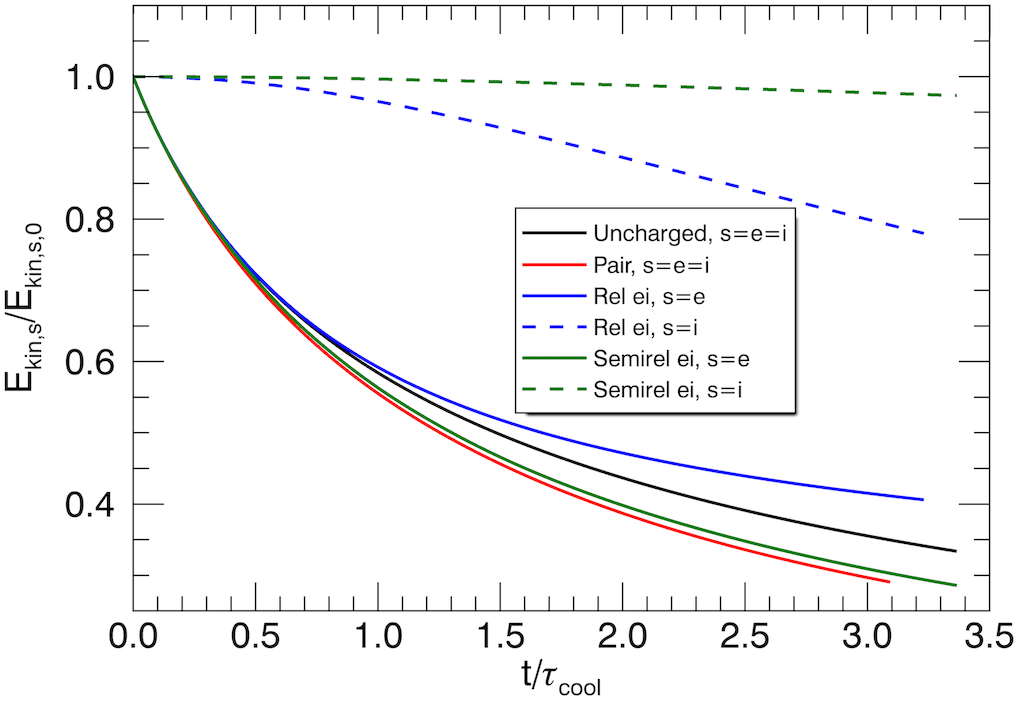}
   \centering
   \caption{\label{fig:evolution} Top panel: evolution of magnetic energy $E_{\rm mag}$ in fluctuations (relative to energy in the mean field $E_{\rm mag, mean}$), in the fiducial simulations (same compositions as Fig.~\ref{fig:brazil}). Also shown are instability thresholds with $C_{\rm thr} = 0.7$ (dotted) and $C_{\rm thr} = 1.4$ (dashed). Middle panel: analytical fits [black, Eq.~\eqref{eq:fit}] to the noise-subtracted magnetic energy $E_{\rm mag}-E_{\rm noise}$ during the linear stage for larger simulations. Bottom panel: evolution of the kinetic energy $E_{{\rm kin},s}$ for electrons (solid) and ions (dashed) in the fiducial simulations.}
 \end{figure}

The onset of the SFHI is associated with a rapid increase in the magnetic energy of the perturbations (from the initial noise floor), as shown in the top panel of Fig.~\ref{fig:evolution} for all three charged cases. The instability starts to grow roughly at the time $\tau_{\rm onset}$ predicted by the linear theory, except for the semirelativistic case, where there is a substantial delay (possibly indicating an overshoot, thus requiring smaller $\tau_{\rm gyro}/\tau_{\rm cool}$ to get into the asymptotic regime).

The initial SFHI growth can be studied in greater detail by subtracting off the contribution to magnetic energy from the PIC noise (caused by the finite number of particles per cell), which we denote~$E_{\rm noise}$. In each case, we define $E_{\rm noise}$ as the mean magnetic energy during an interval of duration $0.015 \tau_{\rm cool}$ shortly before the instability starts to grow. Physically, $E_{\rm noise}$ is associated with fluctuations over a broad range of wavenumbers, as seen from the flat spectrum during the earliest (pre-instability) time in Fig.~\ref{fig:spec_mag}; $E_{\rm noise}$ overwhelms the early-time signal of the instability in~$E_{\rm mag}$, even if the instability dominates the energy at the most unstable wavenumber. Subtracting $E_{\rm noise}$ from $E_{\rm mag}$ allows the initial SFHI growth to be measured over several orders of magnitude in energy, as shown in the middle panel of Fig.~\ref{fig:evolution} (note that this plot uses the larger-box simulations for clarity). We find that this growth can be represented very well by either an exponential or a super-exponential function, as predicted by linear theory in Sec.~\ref{sec:growth}. Motivated by Eq.~\eqref{eq:superexp} describing unregulated growth, we choose to fit the data with a function of the form
\begin{equation}
    \frac{E_{\rm mag}(t)}{E_{\rm mag,mean}} \sim \eta \exp{\left[ \left( \frac{t - \tau_0}{\tau_1} \right)^{3/2} \right]} , \label{eq:fit}
\end{equation}
where $\eta$, $\tau_0$, and $\tau_1$ are fitting parameters representing the amplitude, onset time, and growth time, respectively. Fits of this form are shown for each case by black dashed/dotted lines in the middle panel of Fig.~\ref{fig:evolution}. For all cases, we choose $\eta = 2\times10^{-6}$, representing the level of the seed fluctuations from the noise. Fastest onset and growth occurs for the pair-plasma case, with $\tau_0 = 0.065 \tau_{\rm cool} \approx \tau_{\rm onset, pair}$ [where $\tau_{\rm onset, pair}$ is the theoretical linear onset time from Eq.~\eqref{eq:tauonset} with $C_{\rm thr} = 0.7$ and $\beta_{e0}=20$] and $\tau_1 \approx 9 \times 10^{-3} \tau_{\rm cool}$. This is in close agreement with the linear theory [Eqs.~\eqref{eq:gamma_theory} and \eqref{eq:superexp}], which predicts that $\tau_1 = 2^{-2/3} \tau_{\rm gr} = 7 \times 10^{-3} \tau_{\rm cool}$ for the given plasma parameters and assuming $k\rho_{e0} = 0.4$. The next fastest growth is for the relativistic electron-ion case, with $\tau_0 = 0.11 \tau_{\rm cool} \approx 0.85 \tau_{\rm onset, ei}$ (where $\tau_{\rm onset, ei}$ is the linear onset time with $C_{\rm thr} = 1.4$ and $\beta_{e0}=20$) and $\tau_1 = 1.6 \times 10^{-2} \tau_{\rm cool}$, such that the growth rate is nearly a factor of 2 lower than in the pair case. For comparison, the linear theory calculation predicts $\tau_1 = 9 \times 10^{-3} \tau_{\rm cool}$ in this case (again assuming $k\rho_{e0}=0.4$), somewhat faster than measured. Finally, the semirelativistic plasma has the slowest growth, with $\tau_0 = 0.21 \tau_{\rm cool} \approx 1.6 \tau_{\rm onset, ei}$ and $\tau_1 \approx 2.6 \times 10^{-2} \tau_{\rm cool}$; incidentally, this growth rate is approximately a factor of 3 lower than the pair case. The growth rate is significantly below the linear prediction, $\tau_1 = 1.8 \times 10^{-2} \tau_{\rm cool}$. In summary, the hierarchy of the growth rates is qualitatively consistent with the estimates from linear theory, with pair plasma having the closest agreement with theory. Deviations from linear theory for the electron-ion cases may be due to the non-asymptotic nature of the simulations, as well as corrections from higher-order kinetic terms and the oblique nature of the dominant modes.

Following the initial super-exponential growth, the magnetic energy saturates and then decays. On a much longer timescale, however, the magnetic energy undergoes a nonlinear cyclical evolution of growth and decay (as seen in the top panel of Fig.~\ref{fig:evolution}). The period of the cycles is fastest for the pair-plasma case (having duration $\tau_{\rm cycle} \approx \tau_{\rm cool}/12$) and slowest for the case with sub-relativistic ions (with $\tau_{\rm cycle} \approx \tau_{\rm cool}/3$). Each burst has a peak magnetic energy that is a small fraction of the energy in the mean magnetic field, $\lesssim 0.02 E_{\rm mag, mean}$, and rises above the local minima (between peaks) by a factor of $\sim 3$. The peak amplitudes generally decay over an even longer timescale (of order ${\sim}\tau_{\rm cool}$). The peak amplitude appears to be sensitive to parameters that are not the focus of our study, such as $\tau_{\rm cool}/\tau_{\rm gyro}$ (not shown). Eventually, because of the continuous decrease of $\beta_{\parallel,e}$ from cooling, the plasma will exit the firehose-unstable region of parameter space (and presumably return to a stable cooling evolution).

In principle, the SFHI may affect the cooling rate of the plasma. Pitch-angle scattering will mix the weakly cooled, parallel-propagating particles with strongly cooled, perpendicular-propagating particles, thereby enhancing the cooling rate over the predicted stable evolution (additionally, some electron kinetic energy is transferred to firehose magnetic fields, but this has a weaker impact on the cooling rate). An enhancement of the cooling rate (over the uncharged evolution) is indeed observed for the pair plasma and semirelativistic electron-ion cases, as shown in the bottom panel of Fig.~\ref{fig:evolution}, which displays the evolution of the total kinetic energy $E_{{\rm kin},s}$ for each species ($s \in \{ e, i \}$). The relativistic electron-ion case, on the other hand, experiences slower cooling of the electrons than the uncharged simulation. This implies that electrons experience anomalous heating from some source. The only available source that electrons can draw energy from is the ion kinetic energy; indeed, we find that the anomalous electron heating is compensated by a decrease of ion kinetic energy (dashed blue line in Fig.~\ref{fig:evolution}), indicating collisionless thermal coupling between the two species (studied in Sec.~\ref{sec:ions}). The adjustment to the cooling rate caused by the SFHI in all cases is small (approximately ${\sim}10\%$ over the simulated timescales), but in principle may build to a significant temperature difference over time. We also note that there does not appear to be a simple analytical function that describes~$E_{{\rm kin},s}(t)$: power laws and exponential functions provide poor fits to the simulation data.
 
\begin{figure}
   \includegraphics[width=\columnwidth]{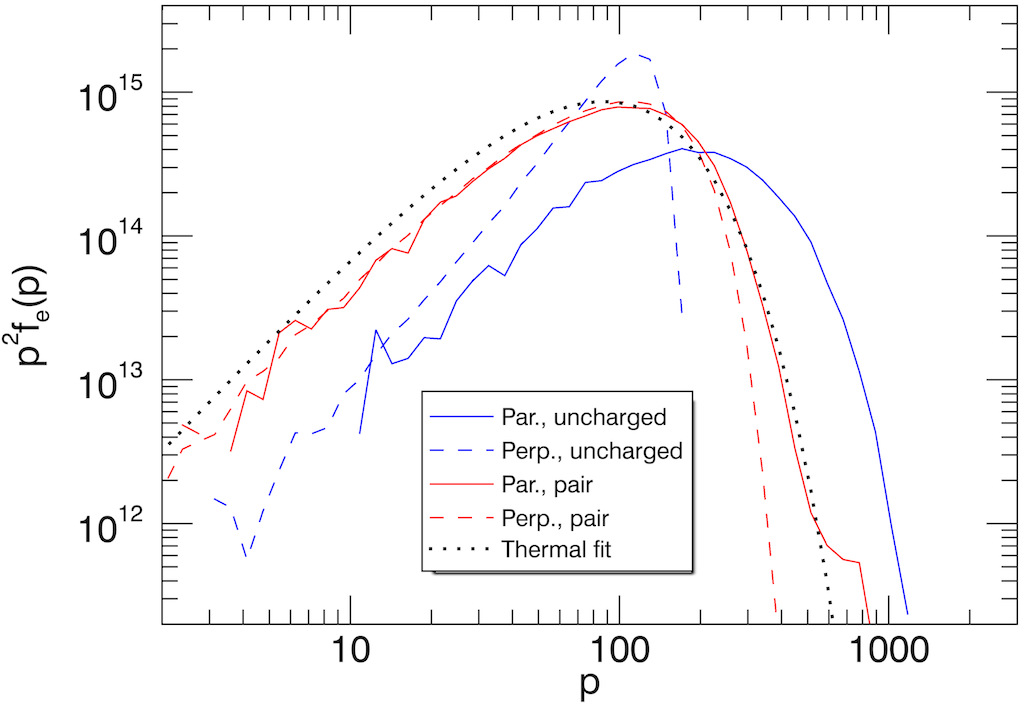}
   \centering
   \caption{\label{fig:dist} Electron momentum distribution in directions parallel (solid) and perpendicular (dashed) to the magnetic field, for the stable uncharged simulation (blue) and the SFHI-unstable pair plasma simulation (red) at $t/\tau_{\rm cool} = 1.4$. A Maxwell--J\"{u}ttner fit is also shown (dotted).}
 \end{figure}

  The electron momentum distribution arising from the SFHI is close to an isotropic thermal distribution, as shown separately for parallel and perpendicular components in Fig.~\ref{fig:dist} (red lines) for the pair-plasma case at $t/\tau_{\rm cool} = 1.4$ (other cases are similar). The stable uncharged simulation, by comparison, has a very strongly nonthermal distribution at this time (matching the analytical solution from Fig.~\ref{fig:analytic}). Thus, the SFHI efficiently thermalizes the distribution by mixing particles with different pitch angles. At large momenta, there is a modest excess of particles in the parallel direction compared to the perpendicular direction, highlighting a persistent asymmetry caused by the continuous synchrotron cooling, of the amount necessary to maintain pressure anisotropy at the marginally unstable state. Although the overall distribution is fairly close to an isotropic Maxwell--J\"{u}ttner distribution, the anisotropy is not represented well by a bi-Maxwell--J\"{u}ttner distribution (i.e., the anisotropies are comparable to the deviations from the Maxwell--J\"{u}ttner distribution in each direction, indicating that the anisotropic component is nonthermal).
  
 \subsection{Ion response and thermal coupling} \label{sec:ions}
 
We next study the effect of the SFHI on ions. In the pair-plasma case, the positrons undergo an evolution that is identical to that of the electrons (as required by symmetry), which was already described in Sec.~\ref{sec:electrons}. Therefore, in this subsection, we consider what happens in the simulations with non-radiative ions.
 
The ions remain in their initial thermal state until the SFHI produces magnetic fields that can perturb them. For the ultra-relativistic electron-ion case, some time after the onset of~SFHI, the ions grow a positive pressure anisotropy, $P_{\perp,i}> P_{\parallel,i}$, as shown in the bottom panel of Fig.~\ref{fig:brazil}. This occurs because the ions interact with the fields produced by the electron firehose.
 
The initial growth of the ion pressure anisotropy can be modeled in the limit of double adiabatic evolution in an increasing magnetic field. In this simplified model, the ions conserve their magnetic moment $\mu = p_\perp^2/2 m_i B$ (associated with periodic Larmor orbit) and parallel momentum $p_\parallel$ (associated with periodic motion along the magnetic field lines) as the firehose fluctuations increase the magnetic-field strength. Under this ``double adiabatic'' evolution, the ion distribution will evolve as
\begin{equation}
f_{i,\rm adi}(p_\perp, p_\parallel,t) = \frac{B_0}{B(t)} f_{i0}\left(\sqrt{\frac{B_0}{B(t)} p_\perp^2 + p_\parallel^2} \right) , \label{eq:adi_dist} 
\end{equation}
where $B(t)$ is the orbit-averaged magnetic field (which we approximate as being the same for all particles) and $f_{i0}(p)$ is the initial isotropic distribution. Assuming a weak fluctuation, $B/B_0 = 1 + \epsilon_{\delta B}$ with $\epsilon_{\delta B} \ll 1$, we can expand the distribution \eqref{eq:adi_dist} and compute pressures to show that the adiabatic evolution follows
\begin{equation}
    \frac{P_\perp}{P_\parallel} \biggr|_{\rm adiabatic} = 1 + \frac{2}{5} \epsilon_{\delta B} = 3 - 2 \frac{\beta_{\parallel}}{\beta_{\parallel,0}} . \label{eq:adi_scaling}
\end{equation}
In deriving Eq.~\eqref{eq:adi_scaling}, we have assumed that ions are ultra-relativistic. The double-adiabatic evolution given by Eq.~\eqref{eq:adi_scaling} is shown by the cyan line in the bottom panel of Fig.~\ref{fig:brazil}. The initial evolution of the simulation is close to this adiabatic prediction, but shifted slightly.

  \begin{figure}
   \includegraphics[width=\columnwidth]{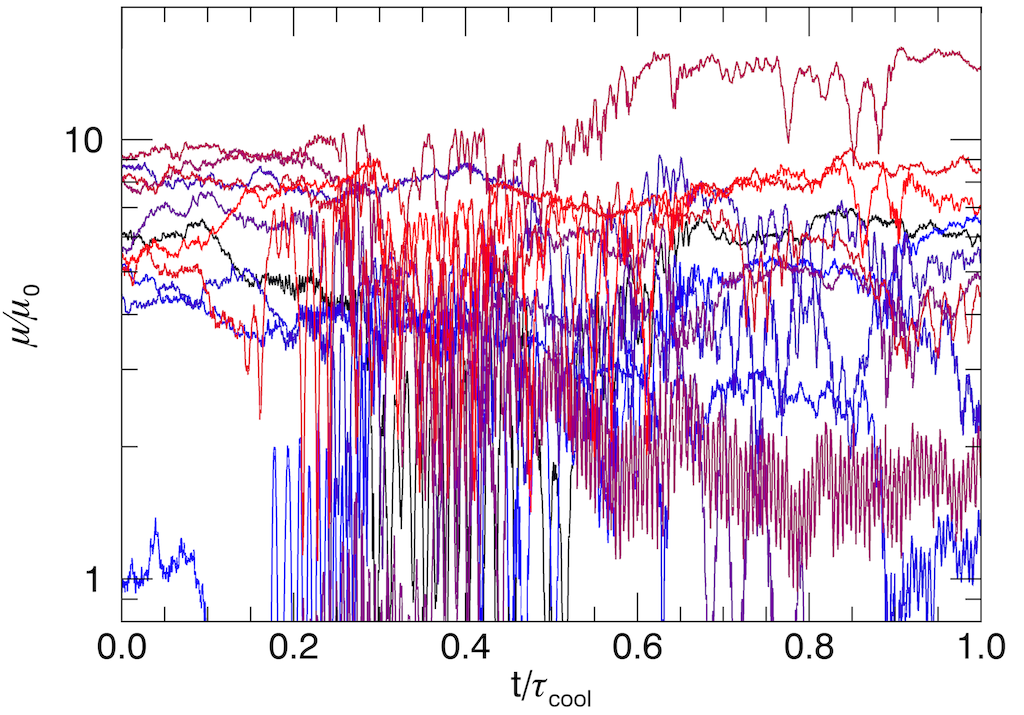} 
   \centering
   \caption{\label{fig:mu} Evolution of the magnetic moment $\mu$ (normalized by $\mu_0 \equiv \overline{p}_0^2/2 m_i B_0$) for a sample of 11 tracked ions in the ultra-relativistic electron-ion case. These particles are chosen randomly subject to the constraint that their initial momentum is larger than $2 \overline{p}_0$.}
 \end{figure}
 
Departures from the double-adiabatic scaling are due to pitch-angle scattering events that break the conservation of $\mu$ and~$p_\parallel$. To demonstrate this, in Fig.~\ref{fig:mu} we show the evolution of the magnetic moment $\mu(t)$ for 11 randomly selected ions (from the ultra-relativistic electron-ion case). To reduce the effect of noise, these ions are chosen with the criterion that their initial momentum is larger than twice the average ion momentum, $p(t=0) > 2 \overline{p}_0$. We find that $\mu$ is well conserved up to a time $t/\tau_{\rm cool} \sim 0.15$ (with small fluctuations due to noise), at which point electron firehose fluctuations emerge. Subsequently, the conservation of $\mu$ is broken for most particles. Note that high-frequency oscillations in $\mu$ are observed for some particles due to their Larmor gyrations; the true adiabatic invariant should be evaluated at the guiding-center position rather than particle position, and would presumably exhibit a smoother evolution.

Returning to the bottom panel of Fig.~\ref{fig:brazil}, once the ion anisotropy becomes sufficiently large, either the ion mirror instability or ion cyclotron instability may be triggered. The threshold for the classical mirror instability can be written as $P_{\perp,i}/P_{\parallel,i} \gtrsim 1/[2(1+\sqrt{1+4/\beta_{\parallel,i}})]$, while the one for the ion cyclotron instability is typically fit by $P_{\perp,i}/P_{\parallel,i} \gtrsim 1 + A/\beta_{\parallel,i}^{b}$ with $A$ a free coefficient and $b \approx 0.4$ \citep[e.g.,][]{gary_lee_1994, hellinger_etal_2006, sironi_narayan_2015}; note that these thresholds do not account for relativistic effects, departures from bi-Maxwellian distributions, and reduced effective (relativistic) mass ratio. We show these thresholds on the bottom panel Fig.~\ref{fig:brazil}, choosing $A = 0.31$ so that the ion-cyclotron threshold fits the trajectory. We find that the ion pressure anisotropy evolves past the mirror threshold and approaches the ion cyclotron threshold before the trajectory is altered (times of $t/\tau_{\rm cool} = 0.4$ and $t/\tau_{\rm cool} = 0.8$ are marked, with the latter corresponding to the time at which the ion pressure anisotropy becomes limited). This suggests that the ion cyclotron instability is the dominant secondary instability in our simulations. Further evidence for the lack of efficient mirror growth comes from the polarization of the modes: we do not observe parallel magnetic-field fluctuations ($\delta B_y$) or density fluctuations that would be expected from the mirror instability. Rather, the polarization of the modes remains similar to the electron firehose modes, as expected for the ion cyclotron instability.

The above suggests that the electron-ion thermal coupling observed in our PIC simulations is caused by the ion cyclotron instability. Free energy for the ion-cyclotron magnetic field fluctuations is provided by the ion internal energy (in the form of ion pressure anisotropy). The dissipation of these fluctuations will cause heating of electrons and ions (partitioned in an undetermined way). The end result of this process is net heating of electrons and cooling of ions.

Throughout the evolution, the ion momentum distribution remains close to a thermal distribution in each direction. This is in contrast with the moderate nonthermal anisotropy for the radiatively cooled electrons.

While ions also attain $P_{\perp,i}/P_{\parallel,i} > 1$ in the semi-relativistic case, the pressure anisotropy does not cross the ion instability thresholds and thus the amount of thermal coupling is negligible in this case. The limited amount of ion pressure anisotropy in this case may be a consequence of the scale mismatch between the ion gyroradius $\rho_i$ and electron gyroradius~$\rho_e$, having $\rho_e/\rho_i \sim (T_e/T_i) \theta_i^{1/2}$ in the semi-relativistic regime. When $\rho_i \gg \rho_e$, the ions are unable to interact strongly with the electron-scale firehose modes (in contrast to the relativistic ion case where $\rho_i \sim \rho_e$). Thus, we conclude that electron-ion thermal coupling is limited to the relativistic ion regime in which $\rho_i \sim \rho_e$.

 \section{Discussion} \label{sec:disc}
 
The results described in this paper are broadly applicable to high-energy astrophysical systems that contain hot (relativistic, high-beta) collisionless plasmas. Examples of such systems are pulsar wind nebulae (e.g., the Crab nebula; relativistic pair plasma), black-hole accretion flows (e.g., around low accretion-rate supermassive black holes in M87 and Sgr~A*; relativistic electrons, sub-relativistic ions), relativistic jets from supermassive black holes in active galactic nuclei (including blazars; uncertain composition), and giant radio lobes (relativistic electrons, sub-relativistic ions). The SFHI may act on a fraction of the cooling time, but is limited to regions where the electron plasma beta is sufficiently high.
 
 In realistic scenarios, heating mechanisms (via dissipation of waves, shocks, magnetic fields, etc.) may counteract radiative cooling, and act as an alternative mechanism of pitch-angle scattering instead of SFHI. In principle, the balance of heating and synchrotron cooling may lead to an anisotropic equilibrium distribution \citep{wentzel_1969, melrose_1970}. Additional modeling will be required to understand the relevance of the SFHI in such environments.
 
 In this work, we considered the SFHI triggered by a gradually cooling distribution; the high initial plasma beta ($\beta_0 \gtrsim 20$) required to encounter the SFHI in this evolutionary track may limit the relevance of this scenario. However, the SFHI may also be triggered if a plasma transitions from low beta to high beta while synchrotron cooling concurrently develops (or maintains) $P_{\perp,e}/P_{\parallel,e} \ll 1$. This situation may occur while the plasma heats or expands \citep[e.g.,][]{bott_etal_2021}. In this alternative scenario, the firehose instability will be approached from a different direction in parameter space than considered here, and may onset at much lower beta ($\beta_{\parallel,e} \gtrsim 1$). While the shape of the distribution as it approaches the firehose threshold will generally be different than that considered in this work, we expect that the subsequent properties of the plasma at marginal stability may be qualitatively similar. In particular, we anticipate that the particle distribution will take a near-thermal form with modest anisotropy near the instability threshold.
 
 The transition of a stable, anisotropic low-beta plasma into the firehose unstable regime may have observable consequences. The redistribution of energetic particles from parallel momenta to perpendicular momenta will cause a rapid increase of the synchrotron radiative power in the corresponding energy band. The onset of the SFHI in a localized region (e.g., a hot spot or interface) might thus lead to an observable flare on timescales faster than the cooling time or macroscopic dynamical times. For example, this scenario may apply to emission from ejecta propagating in black-hole accretion flows, as suggested by general-relativistic magnetohydrodynamic simulations \citep{ripperda_etal_2020, ripperda_etal_2022} as an explanation for near-infrared flares in Sgr~A* observed by the GRAVITY Collaboration and others \citep[e.g.,][]{ponti_etal_2017, baubock_etal_2020}. In this case, the ejecta is produced near the jet-disk interface, and thus will begin in a highly magnetized ($\beta \ll 1$), strongly radiating state where pressure anisotropy will accumulate. As it expands and mixes with the ambient accreting plasma, it may reach $\beta \gtrsim 1$, at which point the SFHI will be triggered. This effect may complement dynamical plasma models for the emission \citep[e.g.,][]{dexter_etal_2020, ball_etal_2021}.
 
Aside from radiative signatures, our work sheds new light on the problem of electron-ion thermal coupling in collisionless plasmas, which has been investigated extensively due to its applications to models of radiatively inefficient accretion flows \citep[e.g.,][]{begelman_chiueh_1988, sironi_narayan_2015, sironi_2015,zhdankin_etal_2019, zhdankin_etal_2021}. The lack of thermal coupling by SFHI in the semirelativistic regime implies that high ion-to-electron temperature ratios may be maintained in astrophysical system such as radiatively inefficient accretion flows, supporting conclusions in our previous work \citep{zhdankin_etal_2021}. However, thermal coupling caused by SFHI in the relativistic ion regime may be relevant to some extreme astrophysical systems such as relativistic jets and gamma-ray bursts.

Our work focused on the regime where electrons are ultra-relativistic. In some systems (e.g., stellar coronae and laboratory experiments), electrons are instead sub-relativistic, in which case cyclotron cooling occurs. Cyclotron cooling reduces perpendicular pressure in a way similar to synchrotron cooling \citep{kennedy_helander_2021a}; however, the Coulomb collision time is typically much faster than the cyclotron cooling time (particularly in high-beta plasmas), which makes it very difficult to develop sufficient anisotropy to trigger the firehose instability. The evolution of a cyclotron-cooling plasma subject to Coulomb collisions has been solved analytically by \cite{kennedy_helander_2021a, kennedy_helander_2021b}.

Finally, one of the key assumptions of our work is that the plasma is optically thin. In the alternative case of an optically thick plasma, additional radiative effects such as radiation pressure, synchrotron self-absorption, and pair creation (in the more extreme regimes) will need to be included. It is unclear whether the firehose instability can be triggered in the presence of such effects.

\section{Conclusions} 
\label{sec:conc}

In this work, we demonstrated that a high-beta, synchrotron-cooling collisionless plasma is unstable to the firehose instability in a process that we called the SFHI. We studied the SFHI analytically (in the linear approximation) and numerically (with 2D PIC simulations). For the onset and linear stage, the PIC simulations validated the linear theory to a good approximation (despite the latter requiring several simplifying assumptions). For the nonlinear stage, the PIC simulations revealed that the SFHI undergoes cyclic evolution near the marginally unstable state. We found that the SFHI will efficiently redistribute the radiating particles in pitch angle, leading to a weaker anisotropy than would be anticipated from stable synchrotron cooling. The SFHI may also cause ions, if relativistic, to cool through collisionless thermal coupling mediated by the secondary ion cyclotron instability (in other regimes, one may speculate that mirror instability can play a similar role). The SFHI will influence the radiative properties of synchrotron-cooled plasmas, by causing bursty emission and altering the cooling rate (albeit only moderately). 

The SFHI-unstable system considered in our work is notable as a simple example of a homogeneous collisionless plasma whose thermal state is out of equilibrium. The spontaneous development of nontrivial dynamics and increasing complexity is possible because the photons act as a sink for energy and entropy in the plasma. This problem may serve as a useful testing ground for studying collisionless relaxation of plasma, a topic that remains poorly understood \citep[e.g.,][]{zhdankin_2022}.

Future work is required to carefully study the dependence of the results on the parameters $\beta_0$ and $\tau_{\rm gyro}/\tau_{\rm cool}$. 3D PIC simulations are necessary to properly study the nonlinear state, energy cascades, and inverse energy transfer. It will also be important to broaden consideration away from homogeneous initial conditions, since the cyclic firehose bursts observed in the present work may become decorrelated in a large inhomogeneous medium. Finally, it will be important to place the SFHI in the context of other radiative kinetic plasma instabilities, which have yet to be explored.

\begin{acknowledgements}
The authors are grateful to Archie Bott, Per Helander, and Bart Ripperda for useful conversations. V.Z.~is supported by a Flatiron Research Fellowship at the Flatiron Institute, Simons Foundation. Research at the Flatiron Institute is supported by the Simons Foundation. This work was performed in part at {bf the} Aspen Center for Physics, which is supported by National Science Foundation grant PHY-1607611. M.W.K.~thanks the Institut de Plan\'etologie et d’Astrophysique de Grenoble (IPAG) for its hospitality and visitor support while this work was in progress. D.A.U. gratefully acknowledges support from NASA grants 80NSSC20K0545 and 80NSSC22K0828 and from NSF grants AST-1806084 and AST-1903335. This work used the Extreme Science and Engineering Discovery Environment (XSEDE), which is supported by National Science Foundation grant number ACI-1548562. This work used the XSEDE supercomputer Stampede2 at the Texas Advanced Computer Center (TACC) through allocation TG-PHY160032 \citep{xsede}.
\end{acknowledgements}

\software{Zeltron \citep{cerutti_etal_2013}}



\appendix

\section{Linear instability derivation} \label{appendix}

In this Appendix, we derive the linear firehose instability criterion for an anisotropic distribution of synchrotron-cooling relativistic electrons (gradually developing from an initial thermal distribution). We consider the case of non-radiating (and thus pressure-isotropic) ions, which may be sub-relativistic or ultra-relativistic, but comment on the case of a pair plasma at the end of the section. Throughout the derivation, we also adopt the assumptions of slow cooling (relative to instability timescale), wavevector parallel to the background magnetic field, and low-frequency/long-wavelength fluctuations. These assumptions must be relaxed for a rigorous prediction of the SFHI onset, but we believe the present derivation is sufficient to illustrate the pertinent aspects of the~SFHI.

We begin with the (relativistic) Vlasov equations for the particle distributions $f_s(\boldsymbol{x},\boldsymbol{p},t)$, where $s \in \{ e, i \}$ denotes the particle species (electron or ion):
\begin{align}
\partial_t f_s + \bb{v}_s \bcdot \grad f_s + q_s \left(\boldsymbol{E} + \boldsymbol{v}_s \btimes \boldsymbol{B} \right)\bcdot \frac{\partial f_s}{\partial \boldsymbol{p}} &= 0 , \label{eq:vlas1}
\end{align}
where $\boldsymbol{v}_s = \boldsymbol{p} / \sqrt{m_s^2 + p^2}$; in this Appendix, units are such that $c = 1$. We have ignored the electron radiation reaction force $\boldsymbol{F}_{\rm sync}$ in Eq.~\eqref{eq:vlas1} under the assumption that cooling is slow compared to the instability (i.e., the cooling timescale is much longer than the inverse of the firehose growth rate). The system is closed with Maxwell's equations for the electric and magnetic fields,
\begin{equation}
\partial_t \boldsymbol{E} = \grad \btimes \boldsymbol{B} - 4\pi \boldsymbol{J}, \qquad 
\partial_t \boldsymbol{B} = - \grad \btimes \boldsymbol{E} , \label{eq:maxo}
\end{equation}
where $\boldsymbol{J} = e \int{\rm d}^3p \, (\boldsymbol{v}_i f_i - \boldsymbol{v}_e f_e)$ is the current density.

In the slow-cooling limit, we can treat the analytical solution for the synchrotron-cooling electron distribution [Eq.~\eqref{eq:solexp} in the main body] at a given time $t = t_0$ as a stationary background, assuming the early-time limit, $t_0/\tau_{\rm cool} \sim \epsilon_1 \ll 1$. The background electron distribution can then be expressed as $f_{e,{\rm b}}(\boldsymbol{p}) = f_{e0}(p) + f_{\rm rad}(p,\theta,t_0)$, where $f_{e0}$ is the initial ultra-relativistic Maxwell--J\"{u}ttner distribution and $f_{\rm rad}$ is the anisotropic component that arises from radiative cooling,
\begin{align}
f_{\rm rad} &=  \left(4-\frac{p}{p_T} \right) \frac{p t_0}{3 p_T \tau_{\rm cool}} \sin^2{\theta} f_{e0}(p) .
\end{align}
Note that the electrons are assumed to be ultra-relativistic, $p \sim p_T \gg m_e$. The background ions are assumed to maintain their initial (isotropic) distribution, $f_{i,{\rm b}}(\boldsymbol{p}) = f_{i0}(p)$.

We consider linear perturbations to the background quantities at times $t = t_0 + \Delta t$, with $\Delta t \ll t_0$ under the assumed orderings. Thus,
\begin{align}
\boldsymbol{B}(\boldsymbol{x},t) &= B_0 \hat{\boldsymbol{z}} + \delta\boldsymbol{B} \,\rme^{\imag k z- \imag \omega \Delta t} \nonumber \\
\boldsymbol{E}(\boldsymbol{x},t) &= \delta\boldsymbol{E} \,\rme^{\imag k z - \imag \omega \Delta t} \nonumber \\
f_e(\boldsymbol{x},\boldsymbol{p},t) &= f_{e,{\rm b}}(\boldsymbol{p}) + \delta f_e(\boldsymbol{p}) \,\rme^{\imag k z- \imag \omega \Delta t} \nonumber \\
f_i(\boldsymbol{x},\boldsymbol{p},t) &= f_{i,{\rm b}}(\boldsymbol{p}) + \delta f_i(\boldsymbol{p}) \,\rme^{\imag k z- \imag \omega \Delta t} ,
\end{align}
with the ordering $\delta B/B_0 \sim \delta f_s/f_{s,{\rm b}} \sim \epsilon_2 \ll 1$, where $\boldsymbol{B}_0 = B_0 \hat{\boldsymbol{z}}$ is the mean magnetic field (note that the coordinate orientation differs from the PIC simulations described in the paper). We have assumed that the wavevector $\boldsymbol{k} = k \hat{\boldsymbol{z}}$ is parallel to~$\boldsymbol{B}_0$, and the mean electric field is zero. To have zero net charge, $\rho_q = \grad \bcdot \boldsymbol{E} / 4\pi = 0$, we require the parallel electric field to vanish, $\delta E_z = 0$; similarly, $\delta B_z = 0$ from $\grad \bcdot \boldsymbol{B} = 0$. 

The perturbed Vlasov--Maxwell equations (to first order in $\epsilon_2$) are
\begin{align}
(-\imag \omega + \imag \boldsymbol{k} \bcdot \boldsymbol{v}_s) \delta f_s + q_s \boldsymbol{v}_s \btimes \boldsymbol{B}_0 \bcdot \frac{\partial \delta f_s}{\partial \boldsymbol{p}} + q_s \delta \boldsymbol{E} \bcdot \frac{\partial f_{s,{\rm b}}}{\partial \boldsymbol{p}} + q_s \boldsymbol{v}_s \btimes \delta \boldsymbol{B} \bcdot \frac{\partial f_{s,{\rm b}}}{\partial \boldsymbol{p}} &= 0 , \nonumber \\
\imag \omega \left( 1 - \frac{k^2}{\omega^2} \right) \delta \boldsymbol{E} &= 4 \pi \boldsymbol{J} , \label{eq:max}
\end{align}
where $\delta \boldsymbol{B} = \boldsymbol{k} \btimes \delta \boldsymbol{E} / \omega$. We will recast the Vlasov equation in spherical momentum coordinates, $\boldsymbol{p} = p( \cos{\phi} \sin{\theta} \hat{\boldsymbol{x}} + \sin{\phi} \sin{\theta} \hat{\boldsymbol{y}} + \cos{\theta} \hat{\boldsymbol{z}})$. Without loss of generality, $\delta \boldsymbol{E} = \delta E_x \hat{\boldsymbol{x}}$, so that $\delta \boldsymbol{B} = \hat{\boldsymbol{y}} k \delta E_x / \omega$. Then the Vlasov part of Eq.~\eqref{eq:max} becomes
\begin{align}
\imag \left(- \omega + \frac{p k \cos{\theta} }{\sqrt{m_s^2 + p^2}} \right) \delta f_s - \frac{q_s B_0}{\sqrt{m_s^2 + p^2}} \frac{\partial \delta f_s}{\partial \phi} + q_s  \delta E_x \cos{\phi} \left( \sin{\theta}  \frac{\partial f_{sb}}{\partial p} + \frac{\cos{\theta} }{p} \frac{\partial f_{sb}}{\partial\theta} - \frac{k/\omega}{\sqrt{m_s^2 + p^2}} \frac{\partial f_{sb}}{\partial \theta} \right) = 0 . \label{eq:lin}
\end{align}
This can be expressed in the form
\begin{align}
\frac{\partial \delta f_s}{\partial \phi} = \imag a_s \delta f_s + b_s \cos{\phi} , \label{eq:fsdiff}
\end{align}
where $a_s$ and $b_s$ are constant with respect to $\phi$.  Specifically, for each species, we have
\begin{align}
a_i &= \frac{1}{e B_0} \left(- \omega \sqrt{m_i^2 + p^2} + p k \cos{\theta} \right) , \qquad
& a_e &= - \frac{1}{e B_0} p \left(- \omega + k \cos{\theta} \right), \nonumber \\
b_i &= \frac{\delta E_x}{B_0} \sin{\theta}  \sqrt{m_i^2 + p^2} \frac{\partial f_{i0}}{\partial p} , \qquad
& b_e &= \frac{\delta E_x}{B_0} \left( p\sin{\theta} \frac{\partial f_{e0}}{\partial p} + p\sin{\theta}  \frac{\partial f_{\rm rad}}{\partial p} + \cos{\theta}  \frac{\partial f_{\rm rad}}{\partial \theta}  - \frac{k}{\omega}  \frac{\partial f_{\rm{rad}}}{\partial \theta}  \right) .
\end{align}
The solution to Eq.~\eqref{eq:fsdiff} is
\begin{align}
\delta f_s &= \frac{\sin{\phi} - \imag a_s \cos{\phi}}{1 - a_s^2} \, b_s + C_0 \,\rme^{\imag a_s \phi} ,
\end{align}
where $C_0$ is a constant of integration. For consistency with periodic boundary conditions in $\phi$, we need $a_s = 2 \pi N$ where $N$ is an integer. This is not satisfied in general, so we set $C_0 = 0$.

Having obtained $\delta f_s$, we must now compute $\boldsymbol{J}$ and insert it into Maxwell's equation to obtain the dispersion relation. The current density for each species, $\boldsymbol{J}_s = q_s \int\rmd^3 p\, v_s (\cos{\phi} \sin{\theta} \hat{\boldsymbol{x}} + \sin{\phi} \sin{\theta} \hat{\boldsymbol{y}} + \cos{\theta} \hat{\boldsymbol{z}}) \delta f_s$, takes the form
\begin{align}
\boldsymbol{J}_s &= q_s \int\rmd p \rmd\theta \rmd\phi \, p^2 \sin{\theta} v_s (\cos{\phi} \sin{\theta} \hat{\boldsymbol{x}} + \sin{\phi} \sin{\theta} \hat{\boldsymbol{y}} + \cos{\theta} \hat{\boldsymbol{z}}) \frac{\sin{\phi} - \imag a_s \cos{\phi}}{1 - a_s^2} \, b_s \nonumber \\
&= \pi q_s \int\rmd p \rmd\theta \,\frac{p^2 b_s v_s \sin^2{\theta}}{1 - a_s^2}  \left( - \imag a_s \hat{\boldsymbol{x}} + \hat{\boldsymbol{y}} \right) . \label{eq:jpreexp}
\end{align}
To make the integrals analytically tractable, we must now resort to the {\it long-wavelength, low-frequency limit}, i.e., the ordering $k \rho_i \sim \omega/\Omega_i \sim a_s \sim \epsilon_3 \ll 1$ (where $\rho_i$ is the ion Larmor radius and $\Omega_i$ is the ion cyclotron frequency). This approximation allows us to expand $(1 - a_s^2)^{-1} \sim 1 + a_s^2 + \cdots$. Computing the integrals in Eq.~\eqref{eq:jpreexp} to first order in $\epsilon_3$ for each species then leads to
\begin{align}
\boldsymbol{J}_i &= - \hat{\boldsymbol{x}} \, \frac{\delta E_x}{B_0} \frac{\imag\omega}{B_0} \int\rmd^3 p\, f_{i0} \frac{(4/3) p^2 + m_i^2}{\sqrt{p^2 + m^2}} - \hat{\boldsymbol{y}} \, e n_0 \frac{\delta E_x}{B_0} ,\nonumber \\
\boldsymbol{J}_e &= - \hat{\boldsymbol{x}} \, \frac{\delta E_x}{B_0} \frac{\imag\omega}{B_0} 4 n_0 p_T \left[ 1 - \frac{2}{15} \frac{t_0}{\tau_{\rm cool}} \left( 7 - \frac{k^2}{\omega^2} \right) \right] + \hat{\boldsymbol{y}}\, e n_0 \frac{\delta E_x}{B_0} .
\end{align}
The $\hat{\boldsymbol{y}}$ component of the total current $\boldsymbol{J} = \boldsymbol{J}_e + \boldsymbol{J}_i$ is therefore zero, while the $\hat{\boldsymbol{x}}$ component is
\begin{align}
J_x = - \frac{\imag \omega}{B_0^2} \left[  \int\rmd^3 p\, \frac{(4/3) p^2 + m_i^2}{\sqrt{p^2 + m_i^2}} f_{i0} + 4 n_0 p_T - \frac{8}{15} \frac{t_0}{\tau_{\rm cool}} \left( 7 - \frac{k^2}{\omega^2} \right)  n_0 p_T  \right] \delta E_x . \label{eq:jx:gen}
\end{align}
To complete the remaining integral over the ion momentum, we need to consider the ultra-relativistic and non-relativistic limits for ions separately.

{\bf Relativistic ions:} First consider ultra-relativistic ions ($\theta_{i0} \equiv T_{i0}/m_i \gg 1$) with a Maxwell--J\"{u}ttner distribution,
\begin{align}
f_{i0}(p)  &=\frac{n_0}{8\pi T_{i0}^3} \,\rme^{-p/T_{i0}} .
\end{align}
In this case, Eq.~\eqref{eq:jx:gen} becomes
\begin{align}
J_x = - \frac{4 \imag \omega n_0}{B_0^2} \left[ T_{i0} + T_{e0} - \frac{2}{15} \frac{t_0}{\tau_{\rm cool}} \left( 7 - \frac{k^2}{\omega^2} \right) T_{e0} \right] \delta E_x ,
\end{align}
where $T_{e0} = p_T$. Combining with Maxwell's equation [Eq.~\eqref{eq:max}] and assuming $T_{i0} = T_{e0} = T_0$, we arrive at
\begin{align}
 \frac{\omega^2}{k^2} &= \frac{1 - (2/15) \beta_0 t_0/\tau_{\rm cool}}{1 + 2 \beta_0 - (14/15) \beta_0 t_0/\tau_{\rm cool}} , \label{eq:omegaurei}
\end{align}
where $\beta_0 \equiv 16 \pi n_0 T_0/B_0^2$. Instability occurs when $\omega^2 < 0$, corresponding to sufficiently late times when the numerator is negative,
\begin{align}
\frac{t_0}{\tau_{\rm cool}} > \frac{\tau_{\rm onset, ei}}{\tau_{\rm cool}} \equiv \frac{15}{2} \beta_0^{-1} . \label{eq:instability}
\end{align}
If $\beta_0$ is of order unity, the early-time approximation ($t_0/\tau_{\rm cool} \ll 1$) will break down before the instability occurs. Thus, {\it this linear calculation is only valid for $\beta_0 \gg 1$}. In this limit, the denominator of the right-hand side of Eq.~\eqref{eq:omegaurei} is always positive.

We can compare Eq.~\eqref{eq:instability} with the standard firehose threshold by reframing it in terms of the instantaneous $\beta_\parallel$ and~$P_\perp/P_\parallel$. From Eq.~\eqref{eq:pressures},
\begin{equation}
\frac{t_0}{\tau_{\rm cool}} \approx \frac{15}{8} \left( 1 - \frac{P_{\perp,e}}{P_{\parallel,e}} \right), \qquad 
\beta_0 = 2 \beta_{\parallel,e} \frac{P_{0,e}}{P_{\parallel,e}} \approx \frac{2 \beta_{\parallel,e}}{1 - (8/15)t_0/\tau_{\rm cool}} = 2 \beta_{\parallel,e} \frac{P_{\parallel,e}}{P_{\perp,e}} ,
\end{equation}
and so the threshold becomes
\begin{align}
\frac{P_{\perp,e}}{P_{\parallel,e}} &< \frac{1}{1 + 2/\beta_{\parallel,e}} \approx 1 - \frac{2}{\beta_{\parallel,e}} . \label{eq:fh}
\end{align}
Therefore, in the required limit of validity $\beta_{\parallel,e} \gg 1$, this is consistent with the standard criterion for the relativistic fluid firehose \citep{barnes_scargle_1973}. 

Finally, we note that for short times $\delta t = t_0 - \tau_{\rm onset}$ after the instability begins to grow, we can express the firehose growth rate $\gamma_{\rm f} = \imag \omega$ from Eq.~\eqref{eq:omegaurei}, in the limit of $\beta_0 \gg 1$, as
\begin{align}
    \frac{\gamma_{\rm f}}{k} \sim \left( \frac{1}{15} \frac{\delta t}{\tau_{\rm cool}} \right)^{1/2} . \label{eq:gammaurei}
\end{align}

{\bf Sub-relativistic ions:} Now consider the opposite case of non-relativistic ions ($\theta_{i0} = T_{i0}/m_i \ll 1$, but finite) with a Maxwellian distribution,
\begin{align}
f_{i0}(p) &= \frac{n_0}{(2\pi m_i T_{i0})^{3/2}} \,\rme^{-p^2/2 m_i T_{i0}} .
\end{align}
In this case, we can approximate $[(4/3) p^2 + m_i^2]/(p^2 + m_i^2)^{1/2} \sim m_i + (5/6) p^2/m_i$ since $p/m_i \ll 1$. Then Eq.~\eqref{eq:jx:gen} becomes
\begin{align}
J_x = - \frac{\imag \omega n_0}{B_0^2} \left[ m_i + \frac{5}{2} T_{i0} + 4 T_{e0} - \frac{8}{15} \frac{t_0}{\tau_{\rm cool}} \left( 7 - \frac{k^2}{\omega^2} \right) T_{e0} \right] \delta E_x .
\end{align}
Note that to get any nontrivial effects, we need to keep $T_{e0}/m_i$ finite. Assuming $T_{i0} = T_{e0} = T_0$, combining with Maxwell's equation (Eq.~\eqref{eq:max}) and rearranging, we arrive at the dispersion relation
\begin{align}
\frac{\omega^2}{k^2} &= \frac{1- (2/15) \beta_0 t_0/\tau_{\rm cool}}{1 + v_{\rm A}^{-2} + (13/8) \beta_0 - (14/15) \beta_0 t_0/\tau_{\rm cool}} ,
\end{align}
where $v_{\rm A}^2 = B_0^2/(4\pi m_i n_0)$ is the square of the Alfv\'{e}n speed in the non-relativistic limit. Thus, the numerator becomes negative when $t/\tau_{\rm cool} > 15/(2\beta_0)$, exactly the same as the case of ultra-relativistic ions. The denominator is always positive in the regime of validity.

Although the onset criterion is the same as the ultra-relativistic ion case, the growth rate $\gamma_{\rm f}$ differs. For short times after the instability initiates, we can express $\gamma_{\rm f}$ for the semirelativistic case as
\begin{align}
    \frac{\gamma_{\rm f}}{k} \sim \left( \frac{8 \theta_{i0}}{15} \frac{\delta t}{\tau_{\rm cool}} \right)^{1/2} , \label{eq:gammasrei}
\end{align}
where we assumed that $\beta_0 \gg 1$ and $\theta_{i0} \ll 2/13$ to simplify the expression.

{\bf Pair plasma:}  For a pair plasma, both species cool via synchrotron radiation. It is a straightforward exercise to redo the calculation described in this Appendix with radiating positrons substituting for the non-radiative relativistic ions. The main difference is that the anisotropy from both species contributes to $\boldsymbol{J}$, rather than only that of the electrons. The dispersion relation then reads
\begin{align}
 \frac{\omega^2}{k^2} &= \frac{1 - (4/15) \beta_0 t_0/\tau_{\rm cool}}{1 + 2 \beta_0 - (28/15) \beta_0 t_0/\tau_{\rm cool}} ,
\end{align}
and the instability condition is
\begin{align}
\frac{P_{\perp,e}}{P_{\parallel,e}}  &<  1 - \frac{1}{\beta_{\parallel,e}} . \label{eq:fhpair}
\end{align}
The resulting onset time $\tau_{\rm onset,pair} = 15/(4 \beta_0)$. For short times after the instability begins, the growth rate $\gamma_{\rm f}$ will increase as
\begin{align}
    \frac{\gamma_{\rm f}}{k} \sim \left( \frac{2}{15} \frac{\delta t}{\tau_{\rm cool}} \right)^{1/2} . \label{eq:gammapair}
\end{align}

\bibliography{sfhi_arxiv}{}

\begin{thebibliography}{}
\expandafter\ifx\csname natexlab\endcsname\relax\def\natexlab#1{#1}\fi
\providecommand{\url}[1]{\href{#1}{#1}}
\providecommand{\dodoi}[1]{doi:~\href{http://doi.org/#1}{\nolinkurl{#1}}}
\providecommand{\doeprint}[1]{\href{http://ascl.net/#1}{\nolinkurl{http://ascl.net/#1}}}
\providecommand{\doarXiv}[1]{\href{https://arxiv.org/abs/#1}{\nolinkurl{https://arxiv.org/abs/#1}}}

\bibitem[{Ball {et~al.}(2021)Ball, {\"O}zel, Christian, Chan, \&
  Psaltis}]{ball_etal_2021}
Ball, D., {\"O}zel, F., Christian, P., Chan, C.-K., \& Psaltis, D. 2021, The
  Astrophysical Journal, 917, 8

\bibitem[{Barnes \& Scargle(1973)}]{barnes_scargle_1973}
Barnes, A., \& Scargle, J.~D. 1973, The Astrophysical Journal, 184, 251

\bibitem[{Baub{\"o}ck {et~al.}(2020)Baub{\"o}ck, Dexter, Abuter, Amorim,
  Berger, Bonnet, Brandner, Cl{\'e}net, Du~Foresto, de~Zeeuw,
  {et~al.}}]{baubock_etal_2020}
Baub{\"o}ck, M., Dexter, J., Abuter, R., {et~al.} 2020, Astronomy \&
  Astrophysics, 635, A143

\bibitem[{Begelman \& Chiueh(1988)}]{begelman_chiueh_1988}
Begelman, M.~C., \& Chiueh, T. 1988, The Astrophysical Journal, 332, 872

\bibitem[{Bott {et~al.}(2021)Bott, Arzamasskiy, Kunz, Quataert, \&
  Squire}]{bott_etal_2021}
Bott, A., Arzamasskiy, L., Kunz, M., Quataert, E., \& Squire, J. 2021, The
  Astrophysical Journal Letters, 922, L35

\bibitem[{Cerutti {et~al.}(2013)Cerutti, Werner, Uzdensky, \&
  Begelman}]{cerutti_etal_2013}
Cerutti, B., Werner, G.~R., Uzdensky, D.~A., \& Begelman, M.~C. 2013, The
  Astrophysical Journal, 770, 147

\bibitem[{{Chandrasekhar} {et~al.}(1958){Chandrasekhar}, {Kaufman}, \&
  {Watson}}]{chandrasekhar_1958}
{Chandrasekhar}, S., {Kaufman}, A.~N., \& {Watson}, K.~M. 1958, Proceedings of
  the Royal Society of London A, 245, 435

\bibitem[{Dexter {et~al.}(2020)Dexter, Tchekhovskoy, Jim{\'e}nez-Rosales,
  Ressler, Baub{\"o}ck, Dallilar, De~Zeeuw, Eisenhauer, Von~Fellenberg, Gao,
  {et~al.}}]{dexter_etal_2020}
Dexter, J., Tchekhovskoy, A., Jim{\'e}nez-Rosales, A., {et~al.} 2020, Monthly
  Notices of the Royal Astronomical Society, 497, 4999

\bibitem[{Gary \& Lee(1994)}]{gary_lee_1994}
Gary, S.~P., \& Lee, M.~A. 1994, Journal of Geophysical Research: Space
  Physics, 99, 11297

\bibitem[{Hellinger \& Matsumoto(2000)}]{hellinger_matsumoto_2000}
Hellinger, P., \& Matsumoto, H. 2000, Journal of Geophysical Research: Space
  Physics, 105, 10519

\bibitem[{Hellinger {et~al.}(2006)Hellinger, Tr{\'a}vn{\'\i}{\v{c}}ek, Kasper,
  \& Lazarus}]{hellinger_etal_2006}
Hellinger, P., Tr{\'a}vn{\'\i}{\v{c}}ek, P., Kasper, J.~C., \& Lazarus, A.~J.
  2006, Geophysical research letters, 33

\bibitem[{{Hellinger} \& {Tr{\'a}vn{\'{\i}}{\v
  c}ek}(2015)}]{hellinger_etal_2015}
{Hellinger}, P., \& {Tr{\'a}vn{\'{\i}}{\v c}ek}, P.~M. 2015, Journal of Plasma
  Physics, 81, 305810103

\bibitem[{Kennedy \& Helander(2021{\natexlab{a}})}]{kennedy_helander_2021a}
Kennedy, D., \& Helander, P. 2021{\natexlab{a}}, Journal of Plasma Physics, 87

\bibitem[{Kennedy \& Helander(2021{\natexlab{b}})}]{kennedy_helander_2021b}
---. 2021{\natexlab{b}}, Journal of Plasma Physics, 87

\bibitem[{Kunz {et~al.}(2014)Kunz, Schekochihin, \& Stone}]{kunz_etal_2014}
Kunz, M.~W., Schekochihin, A.~A., \& Stone, J.~M. 2014, Physical Review
  Letters, 112, 205003

\bibitem[{Ley {et~al.}(2022)Ley, Zweibel, Riquelme, Sironi, Miller, \&
  Tran}]{ley_etal_2022}
Ley, F., Zweibel, E.~G., Riquelme, M., {et~al.} 2022, arXiv preprint
  arXiv:2209.00019

\bibitem[{Li \& Habbal(2000)}]{li_habbal_2000}
Li, X., \& Habbal, S.~R. 2000, Journal of Geophysical Research: Space Physics,
  105, 27377

\bibitem[{Melrose(1970)}]{melrose_1970}
Melrose, D.~B. 1970, Astrophysics and Space Science, 6, 321

\bibitem[{Parker(1958)}]{parker_1958}
Parker, E. 1958, Physical Review, 109, 1874

\bibitem[{Ponti {et~al.}(2017)Ponti, George, Scaringi, Zhang, Jin, Dexter,
  Terrier, Clavel, Degenaar, Eisenhauer, {et~al.}}]{ponti_etal_2017}
Ponti, G., George, E., Scaringi, S., {et~al.} 2017, Monthly Notices of the
  Royal Astronomical Society, 468, 2447

\bibitem[{Ripperda {et~al.}(2020)Ripperda, Bacchini, \&
  Philippov}]{ripperda_etal_2020}
Ripperda, B., Bacchini, F., \& Philippov, A.~A. 2020, The Astrophysical
  Journal, 900, 100

\bibitem[{Ripperda {et~al.}(2022)Ripperda, Liska, Chatterjee, Musoke,
  Philippov, Markoff, Tchekhovskoy, \& Younsi}]{ripperda_etal_2022}
Ripperda, B., Liska, M., Chatterjee, K., {et~al.} 2022, The Astrophysical
  Journal Letters, 924, L32

\bibitem[{Rosin {et~al.}(2011)Rosin, Schekochihin, Rincon, \&
  Cowley}]{rosin_etal_2011}
Rosin, M., Schekochihin, A., Rincon, F., \& Cowley, S. 2011, Monthly Notices of
  the Royal Astronomical Society, 413, 7

\bibitem[{Rybicki \& Lightman(1991)}]{rybicki_lightman_1991}
Rybicki, G.~B., \& Lightman, A.~P. 1991, Radiative processes in astrophysics
  (John Wiley \& Sons)

\bibitem[{Schekochihin {et~al.}(2010)Schekochihin, Cowley, Rincon, \&
  Rosin}]{schekochihin_etal_2010}
Schekochihin, A., Cowley, S., Rincon, F., \& Rosin, M. 2010, Monthly Notices of
  the Royal Astronomical Society, 405, 291

\bibitem[{Sironi(2015)}]{sironi_2015}
Sironi, L. 2015, The Astrophysical Journal, 800, 89

\bibitem[{Sironi \& Narayan(2015)}]{sironi_narayan_2015}
Sironi, L., \& Narayan, R. 2015, The Astrophysical Journal, 800, 88

\bibitem[{Stepney(1983)}]{stepney_1983}
Stepney, S. 1983, Monthly Notices of the Royal Astronomical Society, 202, 467

\bibitem[{Towns {et~al.}(2014)Towns, Cockerill, Dahan, Foster, Gaither,
  Grimshaw, Hazlewood, Lathrop, Lifka, Peterson, Roskies, Scott, \&
  Wilkins-Diehr}]{xsede}
Towns, J., Cockerill, T., Dahan, M., {et~al.} 2014, Computing in Science \&
  Engineering, 16, 62

\bibitem[{{Vedenov} \& {Sagdeev}(1958)}]{vedenov_1958}
{Vedenov}, A.~A., \& {Sagdeev}, R.~Z. 1958, Soviet Physics Doklady, 3, 278

\bibitem[{Wentzel(1969)}]{wentzel_1969}
Wentzel, D.~G. 1969, The Astrophysical Journal, 157, 545

\bibitem[{{Werner} {et~al.}(2018){Werner}, {Uzdensky}, {Begelman}, {Cerutti},
  \& {Nalewajko}}]{werner_etal_2018}
{Werner}, G.~R., {Uzdensky}, D.~A., {Begelman}, M.~C., {Cerutti}, B., \&
  {Nalewajko}, K. 2018, Monthly Notices of the Royal Astronomical Society, 473,
  4840

\bibitem[{Zhdankin(2022)}]{zhdankin_2022}
Zhdankin, V. 2022, Physical Review X, 12, 031011

\bibitem[{Zhdankin {et~al.}(2021)Zhdankin, Uzdensky, \&
  Kunz}]{zhdankin_etal_2021}
Zhdankin, V., Uzdensky, D.~A., \& Kunz, M.~W. 2021, The Astrophysical Journal,
  908, 71

\bibitem[{Zhdankin {et~al.}(2019)Zhdankin, Uzdensky, Werner, \&
  Begelman}]{zhdankin_etal_2019}
Zhdankin, V., Uzdensky, D.~A., Werner, G.~R., \& Begelman, M.~C. 2019, Physical
  review letters, 122, 055101

\bibitem[{Zhou {et~al.}(2022)Zhou, Zhdankin, Kunz, Loureiro, \&
  Uzdensky}]{zhou_etal_2022}
Zhou, M., Zhdankin, V., Kunz, M.~W., Loureiro, N.~F., \& Uzdensky, D.~A. 2022,
  Proceedings of the National Academy of Sciences, 119, e2119831119

\end{thebibliography}
\bibliographystyle{aasjournal}

\end{document}